\documentclass[10pt,a4paper]{article}
\usepackage[english]{babel}
\usepackage[latin1]{inputenc}
\usepackage{amsfonts,amsbsy,bm,euscript,mathrsfs}
\usepackage{amssymb,stmaryrd,faktor}
\usepackage[tbtags]{amsmath}
\usepackage{bbm}
\usepackage{graphicx}
\usepackage[title,titletoc]{appendix}
\usepackage[bookmarks=true,colorlinks=true,linkcolor=blue,citecolor=blue,urlcolor=blue,bookmarksnumbered]{hyperref}

\usepackage{dsfont}
\usepackage{collref}
\usepackage{lmodern}
\usepackage{mathrsfs}
\usepackage{mathtools}
\usepackage{bbm}
\usepackage{braket}
\usepackage{slashed}

\usepackage{graphicx}
\usepackage{booktabs}
\usepackage{subfig}
\usepackage{tikz}
\usetikzlibrary{plotmarks,calc,decorations,decorations.pathmorphing}

\numberwithin{equation}{section}

\makeatletter
\renewcommand\section{\@startsection {section}{1}{\z@}
{-3.5ex \@plus -1ex \@minus -.2ex}
{2.3ex \@plus.2ex}
{\normalfont\Large\bfseries}}
\renewcommand\subsection{\@startsection{subsection}{2}{\z@}
{-3.25ex\@plus -1ex \@minus -.2ex}
{1.5ex \@plus.2ex}
{\normalfont\large\bfseries}}
\makeatother

\DeclareMathOperator{\csch}{csch}
\DeclareMathOperator{\sech}{sech}
\def\u{\vec{u}}
\def\v{\vec{v}}

\usepackage{blkarray}

\begin{document}

\thispagestyle{empty}
\begin{flushright}\footnotesize\ttfamily
DMUS-MP-19-10
\end{flushright}
\vspace{2em}

\begin{center}

{\Large\bf \vspace{0.2cm}
{\color{black} \large Norms and scalar products for $AdS_3$}} 
\vspace{1.5cm}

\textrm{Juan Miguel Nieto Garcia\footnote{\texttt{j.nietogarcia@surrey.ac.uk}} and Alessandro Torrielli \footnote{\texttt{a.torrielli@surrey.ac.uk}}}

\vspace{2em}

\vspace{1em}
\begingroup\itshape
Department of Mathematics, University of Surrey, Guildford, GU2 7XH, UK
\par\endgroup

\end{center}

\vspace{2em}

\begin{abstract}\noindent 
We compute scalar products and norms of Bethe vectors in the massless sector of $AdS_3$ integrable superstring theories, by exploiting the general difference form of the $S$-matrix of massless excitations in the pure Ramond-Ramond case, and the difference form valid only in the BMN limit in the mixed-flux case. We obtain determinant-like formulas for the scalar products, generalising a procedure developed in previous literature for standard $R$-matrices to the present non-conventional situation. We verify our expressions against explicit calculations using Bethe vectors for chains of small length, and perform some computer tests of the exact formulas as far as numerical accuracy sustains us. This should be the first step towards the derivation of integrable form-factors and correlation functions for the $AdS_3$ $S$-matrix theory.  

\end{abstract}

\newpage

\overfullrule=0pt
\parskip=2pt
\parindent=12pt
\headheight=0.0in \headsep=0.0in \topmargin=0.0in \oddsidemargin=0in

\vspace{-3cm}
\thispagestyle{empty}
\vspace{-1cm}

\tableofcontents

\setcounter{footnote}{0}

\section{\label{sec:Intro}Introduction}

Integrability in AdS/CFT \cite{Beisert:2010jr,Arutyunov:2009ga} has proceeded to a high level of sophistication, and more and more aspects of the correspondence are now amenable to testing {\it via} integrable methods. This list is currently being populated, in particular, for the version of the correspondence involving superstring theory on $AdS_3\times S^3\times S^3\times S^1$ and $AdS_3\times S^3\times T^4$ \cite{Babichenko:2009dk,rev3,Borsato:2016hud}. The superconformal algebras responsible for the kinematical symmetries are, respectively, the $\mathfrak{D}(2,1;\alpha)\times \mathfrak{D}(2,1;\alpha)$ Lie superalgebra, where $\alpha$ keeps track of the relative size of the radii of the two $S^3$'s, and its In\"on\"u-Wigner contraction for $\alpha \to 0$, namely the superconformal algebra $\mathfrak{psu}(1,1|2)\times\mathfrak{psu}(1,1|2)$. 

The string sigma-model on the backgrounds described above has been demonstrated to be classically integrable in \cite{Babichenko:2009dk,Sundin:2012gc}. The finite-gap equations characterising the semi-classical spectrum have been derived in \cite{OhlssonSax:2011ms,Lloyd:2013wza}. The worldsheet excitations are divided in two classes: massive and massless modes. The massive-massive $S$-matrix has been derived in \cite{Borsato:2012ud,Borsato:2012ss, Borsato:2013qpa,Borsato:2013hoa} by focusing on a residual algebra made out of, respectively, two and four centrally-extended  $\mathfrak{psu}(1|1)$ sub-blocks preserving the BMN vacuum. This construction has been successfully tested against perturbative results, at least for what concerns the massive sector \cite{Sundin:2012gc,Rughoonauth:2012qd,Abbott:2012dd,Beccaria:2012kb,
Beccaria:2012pm,Sundin:2013ypa,Bianchi:2013nra, Bianchi:2013nra1, Bianchi:2013nra2}. The scattering theory for the {\it massless modes} involves more subtle issues which have not yet been fully resolved \cite{Lloyd:2013wza,Sax:2012jv,PerLinus}. The world-sheet analysis including massless modes has been tackled in a series of works \cite{BogdanLatest, Borsato:2014hja, Lloyd, Borsato:2015mma,Abbott:2014rca,MI, Sax:2014mea,Borsato:2016kbm,Borsato:2016xns}. 

The CFT-dual to the $AdS_3$ string theories are more elusive when compared to the traditional five-dimensional version. The first evidence of integrability on the CFT$_2$ side of the correspondence in the case of $AdS_3\times S^3\times T^4$ string theory was found in \cite{Sax:2014mea}, where it was demonstrated how this reproduces the expectations from the early analysis of \cite{Sax:2012jv}. An analysis of BPS operators singled out by the Bethe ansatz against the predictions of a symmetric $T^4$ orbifold analysis was provided in \cite{Baggio:2017kza}. For what concerns instead the $AdS_3\times S^3\times S^3\times S^1$ string theory, a dual CFT proposal has been put forward by \cite{Tong:2014yna}, and a revisited BPS analysis has been performed in \cite{Baggio:2017kza, Eberhardt:2017fsi}, see additional progress obtained in \cite{Gaber1,Gaber2,Gaber3,Gaber4,Gaber5,Gaber6,Gaber7,Eberhardt:2019qcl,GaberdielUltimo}. Further work on this topic can be found in \cite{Borsato:2015mma,Abbott:2013ixa,Sundin:2013uca,Prin, Prin1, Abbott:2015mla,Abbott:2015mla1,Per,Per1,Per2,Per3,Per4,Per5,Per6,Per7,Per8,Per9,Per10,
Pittelli:2014ria,Regelskis:2015xxa,Hoare:2018jim, Pittelli:2017spf}. 

In \cite{DiegoBogdanAle}, the relativistic limit of the basic $AdS_3$ $S$-matrix was discovered to give rise to a non-trivial scattering theory for massless particles, at odds with the traditional behaviour for massive magnons, in the situation where purely left-left and right-right scattering is considered. The results of \cite{DiegoBogdanAle} have shown that this scattering theory of exclusively left-left and right-right moving massless particles in two-dimensions is nothing else than the familiar non-perturbative picture of a critical theory which Zamolodchikov describes {\it via} an integrable massless $S$-matrix. The Thermodynamic Bethe Ansatz (TBA) has been obtained in \cite{DiegoBogdanAle} for such a CFT, and it has been solved exactly for the ground state energy, as it is not an unusual occurrence in purely conformal relativistic situations \cite{Zamol2}. A central charge value of $6$ was obtained, which left room for speculations on what this CFT ought to be. Building on the observation of \cite{AndreaAle} that a new form of exact relativistic invariance, implying difference-form of the $S$-matrix, keeps holding even without taking the BMN limit for massless particles, \cite{Fontanella:2019ury} extended the TBA analysis to the entire massless sector. In the same paper the mixed R-R and NS-NS flux version \cite{Cagnazzo:2012se,Lloyd:2014bsa,OhlssonSax:2018hgc} of the scattering theory, taken in the BMN limit, was shown to develop the same type of phenomena as the ones observed in \cite{DiegoBogdanAle}, and to be amenable to a very similar analysis. There is still to date no analogue for mixed-fluxes of the change of variables to difference-form which \cite{AndreaAle,Fontanella:2019ury} found out works for the massless sector in the pure R-R case away from the BMN limit. 

The CFTs, so far gleaned at in these works solely through their properties emerging from the TBA analysis, require further study to be completely identified. One test they must be subjected to is the profile of form factors \cite{Karo1,Babu,Karo} \cite{Thomas} \cite{masslessform,superMussardo}, as any correlation function can be written as form factor series by introducing the closure relation between each of the local operators appearing in the correlation function. This series proves very useful thanks to the property of form factors of separating dynamical and operator contributions, i.e. they separate the spacetime dependence of the correlation function from the behaviour of the operator. Furthermore, form factors satisfy a set of powerful restrictions \cite{Formfactoraxioms} that, together with some physical intuition from symmetries, are usually enough to completely constrain them. However, the presence of massless excitations creates some problems when studying the thermodynamical limit of such series. The issue originates because form factors involving massless excitations exhibit non-integer power-law behaviour in the volume, making it impossible to take the thermodynamic limit of a finite volume form factor series solely in terms of multiple integrals. For a detailed discussion on the topic and possible methods to circumvent it, see \cite{Kozlowski} and references therein.

In order to compute form factors in an integrable theory we can make use of the quantum inverse scattering problem, which in this context means the reconstruction of local operators in terms of the Bethe operators appearing in the monodromy matrix, to relate them to the computation of scalar products of Bethe vectors. The reconstruction was performed for the inhomogeneous XXZ Heisenberg spin chain in \cite{Kitanine:1999rfm} via factorizing F-matrices and for the homogeneous XYZ spin chain in \cite{Gohmann:1999av} using the properties of the $R$-matrix. These computations were later generalised to any quantum integrable model whose $R$-matrix degenerates into the permutation operator at spectral parameter equal to zero \cite{Maillet:1999re}. Although the $R$-matrices we will be dealing with in this article do not have this property, most of them become the graded permutation operator at zero spectral parameter and a similar procedure can be applied to them.

In this article we focus on the computation of scalar products and norms of off-shell Bethe vectors, i.e. Bethe vectors constructed without restricting their rapidities to fulfil the Bethe equations. Since Korepin \cite{KorepinNorm} and Slavnov \cite{SlavnovScalarProduct} proved respectively that the norm and the scalar product of Bethe vectors in models with $\mathfrak{gl}_2$ symmetry can be expressed as determinants, there has been a tireless effort to find determinant formulas for higher rank algebras and quantum deformations of them. Some examples of these efforts are given by \cite{Res86,Wheeler:2012,Belliard:2012pr,SlavnovGL11,SlavnovGL11Bis,Hutsalyuk:2017tcx,Gromov:2019wmz}. Here we will be using the ``composite model trick'', developed in \cite{PakuliakSlavnov} for computing scalar products in models with $U_q (\mathfrak{gl}_m)$ symmetry, to compute off-shell scalar products in spin chains constructed using the $R$-matrices found in \cite{DiegoBogdanAle} and \cite{Fontanella:2019ury}. In the first case (pure R-R case) the computation is similar to the one we would perform in the trigonometric $\mathfrak{su} (1|1)$ spin chain, but the second case (mixed R-R and NS-NS fluxes) is more subtle. In particular, we will see that the conjugate of the operators constructed from the monodromy matrix cannot be written in terms of the very same operators, but in terms of the operators obtained from a second monodromy matrix defined on a different auxiliary space.

The article is structured as follows. In section 2 we present the necessary details about the spin chains we will analyze. In section 3 we explain the composite model trick and how it applies to the present case. In section 4 we construct the conjugate of the Bethe operators in order to establish which scalar products would give us the correct expression for the norm of Bethe states. In section 5 we construct a determinant expression for the scalar product of off-shell Bethe states constructed using the $R$-matrix described in \cite{DiegoBogdanAle}. In section 6 we construct off-shell scalar products associated to the $R$-matrix described in \cite{Fontanella:2019ury}, both when the dual state is built using the original monodromy matrix and when it is built using the conjugate monodromy matrix. In section 7 we present our conclusions and future perspectives. Appendix A contains the proof of a generalisation of the Cauchy determinant formula to hyperbolic and shifted hyperbolic functions. Appendix B collects some intermediate steps in the computation of the scalar product. Appendix C contains a computation of the norm of on-shell Bethe states using a variation of the composite model trick developed in \cite{JuanMiguel}.

\section{\label{ABA}Recap of Algebraic Bethe Ansatz for $AdS_3$}

\subsection{Pure R-R flux}

In this section we summarise the Algebraic Bethe Ansatz procedure performed in \cite{DiegoBogdanAle}. We start by defining
\begin{equation}
a(\theta) = \mbox{sech} \frac{\theta}{2} \ , \qquad  b(\theta) = \mbox{tanh} \frac{\theta}{2} \ .
\end{equation}
The R matrix we will be dealing with can be rewritten as
\begin{equation}
\label{matrR}
R(\theta) = E_{11} \otimes E_{11} - E_{22} \otimes E_{22} - b(\theta) \, \big( E_{11} \otimes E_{22} - E_{22} \otimes E_{11}\big) - a(\theta) \, \big( E_{12} \otimes E_{21} - E_{21} \otimes E_{12}\big) \ .
\end{equation}
In the above formula we have denoted by $\theta = \theta_1 - \theta_2$ the difference of the suitable variables  utilised for the massless sector of pure R-R $AdS_3$ string theories, cf. comments in the Introduction.

The transfer matrix $\tau$ is defined as the supertrace of the monodromy matrix ${\cal{T}}_0$, namely
\begin{equation}
\tau\big(\theta_0|\vec{\theta} \, \big) = \mbox{str}_0 \, {\cal{T}} \big(\theta_0|\vec{\theta} \, \big) \ , \qquad [{\cal{T}}]_0  \big(\theta_0|\vec{\theta} \, \big) = \prod_{i=1}^L R_{0,i}(\theta_0 - \theta_i) \ , \qquad \prod_{i=1}^L g_i \equiv g_1 ... g_L  \ .
\end{equation}
The auxiliary $0$-th particle has rapidity $\theta_0$, while the {\it quantum} (or {\it frame}) particles have rapidities $\theta_i$, $i=1,...,L$. The rapidities of the quantum particles, which can be interpreted as playing the role of {\it inhomogeneities} on a chain of $L$ sites, are collectively denoted by $\vec{\theta}$. We can rewrite the monodromy matrix as
\begin{equation}
\label{entri}
{\cal{T}}_0  \big(\theta_0|\vec{\theta} \, \big) = E_{11} \otimes A\big(\theta_0|\vec{\theta} \, \big) + E_{12} \otimes B\big(\theta_0|\vec{\theta} \, \big) + E_{21} \otimes C\big(\theta_0|\vec{\theta} \, \big) + E_{22} \otimes D\big(\theta_0|\vec{\theta} \, \big) \ ,
\end{equation}
with $E_{ij}$ being the matrices with all $0$'s but a $1$ in row $i$, column $j$ and living in the $0$ space, and the operators $A,B,C,D$ exclusively acting on the quantum spaces $1,...,L$. 

The Algebraic Bethe Ansatz proceeds by selecting a {\it pseudo-vacuum}, namely a highest-weight eigenvector of the transfer matrix out of which all the magnon-excitations are then created. The pseudo-vacuum of choice here is
\begin{equation}
|0\rangle = \underbrace{|\phi\rangle \otimes ... \otimes |\phi\rangle}_{L \text{ factors}} \ ,
\end{equation} 
such that
\begin{equation}
\tau\big(\theta_0|\vec{\theta} \, \big) |0\rangle = \Lambda_0 \big(\theta_0| \emptyset |\vec{\theta} \, \big) |0\rangle, \qquad \Lambda_0 \big(\theta_0| \emptyset |\vec{\theta} \, \big) = 1 -\prod_{i=1}^L b(\theta_0 - \theta_i) \ . 
\end{equation}
As usual, the pseudo-vacuum is annihilated by the $C$ operation and is an eigenstate of the $A$ and $D$ operators with eigenvalues\footnote{The symbol $\delta$ is defined by (\ref{delt}) and should not be confused with the Dirac delta. We will always use $\delta$ in this sense in the paper.}
\begin{align}
\label{delt}
	A(u ) |0\rangle &=\alpha (u) |0\rangle = |0\rangle \ , & D(u ) |0\rangle &=\delta (u) |0\rangle = \prod_{i=1}^L b(u - \theta_i) |0\rangle \ .
\end{align}
Finally, the $B$ operator can be used as the building block for generic eigenvectors of the transfer matrix with $M$ magnons
\begin{equation}
|u_1,...,u_M \rangle = \prod_{n=1}^M B\big(u_n|\vec{\theta} \, \big) |0\rangle \ .
\end{equation}
Such a multi-magnon state can be seen to be an eigenvector of $\tau$ for arbitrary $M$ by simply relying on the commutation relations of $A$, $B$, $C$ and $D$, which follow from the fundamental RTT relations
\begin{equation}
\label{RTT}
R_{0,0'} (\theta_0 - \theta_0') \, {\cal{T}}_0  \big(\theta_0|\vec{\theta} \, \big) \, {\cal{T}}_{0'} \big(\theta_0'|\vec{\theta} \, \big) \, = \, {\cal{T}}_0  \big(\theta_0'|\vec{\theta} \, \big) \, {\cal{T}}_{0'}  \big(\theta_0|\vec{\theta} \, \big) \, R_{0,0'} (\theta_0 - \theta_0') \ .
\end{equation}
The RTT relations are written for two auxiliary spaces and $L$ quantum spaces. Given that $a(\theta)^2 + b(\theta)^2 = 1$, equation (\ref{RTT}) implies in particular 
\begin{eqnarray}
&&\left[ C\big(\theta_0|\vec{\theta} \, \big) , B\big(\theta_0 '|\vec{\theta} \, \big) \right] = \frac{a(\theta_0-\theta_0')}{b(\theta_0-\theta_0')} \left( D\big(\theta_0'|\vec{\theta} \, \big)A\big(\theta_0|\vec{\theta} \, \big)- D\big(\theta_0|\vec{\theta} \, \big)A\big(\theta_0 '|\vec{\theta} \, \big) \right) \ , \label{comCB} \\
&&A\big(\theta_0|\vec{\theta} \, \big) B\big(\theta_0'|\vec{\theta} \, \big) = \frac{a(\theta_0 - \theta_0')}{b(\theta_0 - \theta_0')} \, B\big(\theta_0|\vec{\theta} \, \big) A\big(\theta_0'|\vec{\theta} \, \big) \, - \, \frac{1}{b(\theta_0 - \theta_0')} \, B\big(\theta_0'|\vec{\theta} \, \big) A\big(\theta_0|\vec{\theta} \, \big) \label{AB} \ ,
\end{eqnarray}
and the same relation where $A$ is replaced by $D$. This allows to commute the transfer matrix
\begin{equation}
\tau \big(\theta_0|\vec{\theta} \, \big) = A \big(\theta_0|\vec{\theta} \, \big) - D \big(\theta_0|\vec{\theta} \, \big)  \ ,
\end{equation}
through the series of $B$'s in the $M$-magnon state, \footnote{We recall that $[B(u_1),B(u_2)]=0=[C(u_1),C(u_2)]$ for any value of the magnon rapidities $u_1$ and $u_2$, where $[.,.]$ denotes the commutator, even though the operators $B$ and $C$ are fermionic.} and gather the eigenvalue
\begin{eqnarray}
\label{eige}
&&\tau\big(\theta_0|\vec{\theta} \, \big) |u_1,...,u_M \rangle = \Lambda_M \big(\theta_0|\u \, |\vec{\theta} \, \big) |u_1,...,u_M \rangle + \sum_{j=1}^M Z_j \big(\theta_0|\u \, |\vec{\theta} \, \big) |u_1,...,u_{j-1}, \theta_0, u_{j+1},...,u_M\rangle  \nonumber\\
&&\Lambda_M \big(\theta_0|\u \, |\vec{\theta} \, \big)= \left[ \alpha (\theta_0 ) -\delta (\theta_0) \right] \prod_{n=1}^M \frac{1}{b(u_n - \theta_0)} = \Big[1 - \prod_{i=1}^L b(\theta_0 - \theta_i)\Big] \prod_{n=1}^M \frac{1}{b(u_n - \theta_0)} \ .
\end{eqnarray}
Since only for a value of the unwanted terms $Z_j$ - whose explicit expression we do not report here - equal to $0$ one has that $|u_1,...,u_M \rangle$ is an eigenstate, and since the set of $Z_j$ collect the contributions from the first term on the r.h.s. of (\ref{AB}), it is possible to see that one can set every $Z_j=0$ by imposing the following set of {\it auxiliary} Bethe equations
\begin{equation}
\label{auxi}
\prod_{i=1}^L b(u_n - \theta_i) = 1 \ , \qquad \forall \, \, n=1,...,M \ .
\end{equation}
The momentum-carrying equations are given by
\begin{equation}
e^{i e^{\theta_k} L} \Lambda_M(\theta_k|\u|\vec{\theta}\, ) = 1, \qquad k=1,...,N \ ,
\end{equation}
which, since $b(0)=0$, reduces to
\begin{eqnarray}
\label{syste}
e^{i L p(\theta_k)} \, \prod_{i=1}^{L} S(\theta_k - \theta_i) \, \prod_{i=1}^M \mbox{coth} \frac{\beta_i - \theta_k}{2} = 1, \qquad k=1,...,L,
\end{eqnarray}
with $p(\theta_k)$ the appropriate momentum parameterisation in terms of difference-form rapidities.
In the last step we have also multiplied the transfer matrix eigenvalue by the appropriate product of dressing factors \cite{DiegoBogdanAle}, {\it i.e.} 
\begin{equation}
\prod_{i=1}^L \Phi(\theta_0 - \theta_i)= \prod_{i=1}^L \prod_{\ell=1}^\infty \frac{\Gamma^2 \left(\ell - \frac{\theta_0 - \theta_i}{2 \pi i} \right) \, \Gamma \left(\ell +\frac{1}{2} + \frac{\theta_0 - \theta_i}{2 \pi i} \right) \,\Gamma \left(\ell -\frac{1}{2} + \frac{\theta_0 - \theta_i}{2 \pi i} \right)}{\Gamma^2 \left(\ell + \frac{\theta_0 - \theta_i}{2 \pi i} \right) \, \Gamma \left(\ell +\frac{1}{2} - \frac{\theta_0 - \theta_i}{2 \pi i} \right) \,\Gamma \left(\ell -\frac{1}{2} - \frac{\theta_0 - \theta_i}{2 \pi i} \right)}  \ .
\end{equation}
The dressing factor will not play any role in the analysis performed this paper, which will mostly be structural and algebraic, and can be included by a redefinition of the functions $\alpha (u) $ and $\delta (u)$.

\subsection{Mixed flux}

In this section we describe the case of the (shifted) massless relativistic limit of the mixed R-R and NS-NS flux $R$-matrix discussed in \cite{Fontanella:2019ury}. The attribute of ``shifted" denotes a very particular shift of the originally massive variables, as explained in that article. We also remind of the close connection this has with the $q$-deformed relativistic $S$-matrix of \cite{Hoare:2011wr} and the Pohlmeyer reduced $S$-matrix of \cite{Hoare:2011fj} and especially of \cite{Ben}, and how this scattering theory does not suffer from lack of unitarity \cite{Timminus1,Tim0,Tim1}- cf. \cite{Fontanella:2019ury} for further details. 

The algebraic Bethe ansatz associated to this $R$-matrix was obtained along very similar lines as the one for pure R-R flux \cite{DiegoBogdanAle}, which was summarised above.
One begins by writing the $R$-matrix in the following form
\begin{multline}
R_{LL}(\theta) = E_{11} \otimes E_{11} + c_{LL}(\theta) E_{22} \otimes E_{22} \\
+ b_{LL}(\theta) (E_{11} \otimes E_{22}+ E_{22} \otimes E_{11}) +a_{LL}(\theta) (E_{21} \otimes E_{12}- E_{12} \otimes E_{21}) \ , \label{RmatrixLL}
\end{multline}
where
\begin{eqnarray}
\label{abc}
&&a_{LL}(\theta) = \frac{e^{\frac{\theta}{2}} (e^{\frac{2 \pi i }{k}} - 1)}{e^{\frac{2 \pi i }{k}+\theta} - 1} \ , \qquad b_{LL}(\theta) = \frac{e^{\frac{\pi i }{k}} (e^\theta - 1)}{e^{\frac{2 \pi i }{k}+\theta} - 1} \ , \qquad
c_{LL}(\theta) = \frac{e^\theta - e^{\frac{2 \pi i }{k}}}{e^{\frac{2 \pi i }{k}+\theta} - 1} \ .
\end{eqnarray}
In the above formula we have denoted by $\theta = \theta_1 - \theta_2$ the difference of the relativistic rapidities valid for the BMN limit of (shifted) mixed-flux $AdS_3$ string theories, cf. comments in the Introduction.
It will prove useful in later calculations to also introduce the function
\begin{equation}
	d (\theta )=e^{-\theta /2} \left( e^{2\pi i /k} -e^\theta \right) \ .
\end{equation}

The transfer matrix once again reads 
\begin{eqnarray}
\tau (\theta_0|\vec{\theta}\, ) = str_0 {\mathcal{T}_{LL}}(\theta_0|\vec{\theta}\, ) \ , \qquad {\left[\cal{T} \right]}_0(\theta_0|\vec{\theta}\, ) = \prod_{i=1}^N R_{0i}(\theta_0 - \theta_i) \ , \nonumber
\end{eqnarray}
with ${\cal{T}}(\theta_0|\vec{\theta}\, )$ being again the monodromy matrix, expressed as a matrix in the auxiliary $0$ space with entries being operators on a chain of frame particles of length $L$, each frame particle having rapidity $\theta_i$.

The RTT relations necessary to perform the Algebraic Bethe Ansatz are
\begin{eqnarray}
\label{rels}
&&A(u_1|\vec{\theta}\, )B(u_2|\vec{\theta}\, ) = X(u_1-u_2) B(u_1|\vec{\theta}\, )A(u_2|\vec{\theta}\, ) +Y(u_1-u_2) B(u_2|\vec{\theta}\, )A(u_1|\vec{\theta}\, ) \ , \nonumber \\
&&D(u_1|\vec{\theta}\, )B(u_2|\vec{\theta}\, ) = X(u_1-u_2) B(u_1|\vec{\theta}\, )D(u_2|\vec{\theta}\, ) + Y(u_1-u_2) B(u_2|\vec{\theta}\, )D(u_1|\vec{\theta}\, ) \ ,
\end{eqnarray}
where the coefficients $X,Y$ are given by
\begin{eqnarray}
\label{xy}
&&X(\theta) = \frac{a_{LL} (\theta)}{b_{LL} (\theta)} = \frac{2 i \, e^{\frac{\theta}{2}} \sin \frac{\pi}{k}}{e^\theta - 1} \ , \qquad  Y(\theta) = \frac{c_{LL} (\theta)}{b_{LL} (\theta)} = \frac{e^{-\frac{i \pi}{k}}(e^\theta - e^{\frac{2 i \pi}{k}})}{e^\theta - 1} \ .
\end{eqnarray}
One should further notice that the $B$ and $C$ operators no longer commute with one another, fulfilling instead
\begin{eqnarray}
&&B(u_1|\vec{\theta}\, ) B(u_2|\vec{\theta}\, ) +c_{LL} (u_1-u_2) B(u_2|\vec{\theta}\, ) B(u_1|\vec{\theta}\, ) =0 \ , \nonumber\\ 
&& C(u_1|\vec{\theta}\, ) C(u_2|\vec{\theta}\, ) + c_{LL} (u_2-u_1) C(u_2|\vec{\theta}\, ) C(u_1|\vec{\theta}\, ) =0 \label{comBandCmixed} \ .
\end{eqnarray}
Another useful commutation relation we will need is
\begin{equation}
	C(u_1|\vec{\theta}\, ) B(u_2|\vec{\theta}\, ) - B(u_2|\vec{\theta}\, ) C(u_1|\vec{\theta}\, )= X(u_1 -u_2) \left[ D(u_2|\vec{\theta}\, ) A(u_1|\vec{\theta}\, ) - D(u_1|\vec{\theta}\, ) A(u_2|\vec{\theta}\, ) \right] \ . \label{CBcommutationLL}
\end{equation}
Now we can proceed to postulate that any eigenstate of the transfer matrix 
\begin{equation}
{\cal{T}}(\theta_0|\vec{\theta}\, ) = A(\theta_0|\vec{\theta}\, ) - D(\theta_0|\vec{\theta}\, ) \ 
\end{equation}
can be built using the same pseudo-vacuum as in the pure R-R case and has the form
\begin{equation}
|u_1,...,u_M\rangle = \frac{B(u_1|\vec{\theta}\, )...B(u_M|\vec{\theta}\, )|0\rangle}{\prod_{i<j} d (u_i - u_j)} \label{symmetrizedstates} \ ,
\end{equation}
provided a set of Bethe-equation conditions are imposed on the rapidities. Here the set of $d$ factors is introduced only to make $|u_1,...,u_M\rangle$ symmetric under the exchange of two rapidities, thanks to the property $d(\theta )=-c_{LL} (\theta) d(-\theta )$. One can easily see that the pseudo-vacuum is itself an eigenstate of the transfer matrix (which one can prove by using a recursive argument)
\begin{equation}
\Big[A(\theta_0|\vec{\theta}\, ) - D(\theta_0|\vec{\theta}\, )\Big] \, |0\rangle =\left[ \alpha_{LL}  (\theta_0 ) -\delta_{LL}  (\theta_0 ) \right] |0\rangle = \Big[1 - \prod_{i=1}^N b_{LL}(\theta_0 - \theta_i)\Big] \, |0\rangle \ . 
\end{equation}


One can then see that $|u_1,...,u_M \rangle$ is an eigenstate of the transfer matrix if we impose the \textit{auxiliary Bethe equations}
\begin{equation}
\label{auxi2}
\prod_{i=1}^N b_{LL} (u_j - \theta_i) = 1, \qquad j=1,...,M \ ,
\end{equation}
which eliminate the unwanted term in the expression\footnote{Notice that the expression for the unwanted term (5.9) in \cite{Fontanella:2019ury} lacks a factor accumulated by the repeated use of (\ref{comBandCmixed}) and related to the fact that the $B$ operators are not symmetric. Thanks to the structure of the unwanted terms, this error did not affect the set of Bethe ansatz equations, and ultimately the normalised Bethe vectors are the same.}
\begin{eqnarray}
\label{use}
&&\Big[A(\theta_0|\vec{\theta}\, ) - D(\theta_0|\vec{\theta}\, )\Big] \, |u_1,...,u_M\rangle = \Lambda_M^{(LL)}(\theta_0|\u|\vec{\theta}\, )\, |u_1,...,u_M\rangle \nonumber\\
&&\, \, \, + \sum_{j=1}^M X(\theta_0 - u_j)\Big[\prod_{k \neq j}^M Y(u_j - u_k)\frac{d(\theta_0-u_k)}{d(u_j-u_k)}\Big] \, \Big[1 - \prod_{i=1}^N b_{LL} (u_j - \theta_i)\Big]|u_1,...,u_{j-1}, \theta_0, u_{j+1},...,u_M\rangle \ .\nonumber
\end{eqnarray}
The eigenvalue of the transfer matrix is then
 \begin{equation}
\Lambda_M^{(LL)}(\theta_0|\u|\vec{\theta}\, ) = \Big[\prod_{i=1}^M Y(\theta_0 - u_i)\Big] \Big[1 - \prod_{s=1}^N b_{LL} (\theta_0 - \theta_s)\Big] \ .
\end{equation}

The \textit{momentum carrying} Bethe equations are then obtained as
\begin{equation}
e^{i e^{\theta_k} L} \Lambda_M^{(LL)}(\theta_k|\u|\vec{\theta}\, ) = 1, \qquad k=1,...,N \ .
\end{equation}
Reinstating the dressing factor - which we denote as $\Phi_{LL}$ and whose details we shall not need here, but are contained in \cite{Fontanella:2019ury} - and noticing that $b_{LL}(0)=0$, we get the following momentum-carrying equation
\begin{equation}
\label{main}
e^{i e^{\theta_k} L} \prod_{j=1}^N \Phi_{LL}(\theta_k - \theta_j)  \prod_{i=1}^M Y(\theta_k - \beta_i) = 1, \qquad k=1,...,N \ ,
\end{equation}
with $Y(\theta)$ given in (\ref{xy}).

So far we have exclusively dealt with the so-called $L$ representation \cite{Fontanella:2019ury}. Later we will want to obtain the Hermitian conjugate of the $B$ operator in the case of mixed flux and it turns out that we now need to resort to a combined approach, where we call into the game the $R$ representation as well - related to the $L$ one by crossing symmetry.

We can construct a monodromy matrix which has the same physical (quantum) space as the previous one, namely all $L$ representations, but with the auxiliary space being in the $R$ representation
\begin{eqnarray}
\tau_{RL}(\theta_0|\vec{\theta}\, ) = str_0 {\cal{T}}_{RL}(\theta_0|\vec{\theta}\, ) \ , \qquad \left[{\cal{T}}_{RL}\right]_0(\theta_0|\vec{\theta}\, ) = \prod_{i=1}^N \frac{[R_{RL}]_{0i}(\theta_0 - \theta_i)}{a_{RL} (\theta_0 - \theta_i)} \label{transfer-monodromy-RL} \ ,\nonumber
\end{eqnarray}
where $a_{RL}$ is a function we will introduce later\footnote{The function $a_{RL}(\theta)$ will have zeros on the imaginary axis for special values of the argument (and at infinity). We can avoid such points for real rapidities, and in general by considering regions away from such special points.}, added to simplify the conjugation of the Bethe operators. We will denote by $\tilde{A}, \tilde{B}$, etc. the operators acting on the quantum space obtained from this monodromy matrix. Furthermore, $\mathcal{T}_{RL}$ satisfies the following RTT relations
\begin{equation}
\left[R_{RR}\right]_{0,0'}(\theta - \theta') \, [\mathcal{T}_{RL}]_0(\theta) \, [\mathcal{T}_{RL}]_{0'}(\theta') = [\mathcal{T}_{RL}]_{0'}(\theta') \, [\mathcal{T}_{RL}]_0(\theta) \, \left[R_{RR}\right]_{0,0'}(\theta - \theta') \ ,  
\end{equation}
which translate into commutation relations equivalent to (\ref{rels}) and (\ref{comBandCmixed}). In addition to that, the vacuum we found for $\mathcal{T}_{LL}$ is also an eigenstate of $\tilde{A}$ and $\tilde{D}$ with eigenvalues
\begin{eqnarray}
\tilde{A} (\theta_0 | \vec{\theta} ) |0\rangle=|0\rangle  \ , \qquad \tilde{D} (\theta_0 | \vec{\theta} ) |0\rangle=\Delta (\theta_0) |0\rangle= \left[ \delta (\theta_0^*) \right]^* |0\rangle \ .\nonumber
\end{eqnarray}
These statements together imply that both monodromy matrices, $\mathcal{T}_{LL}$ and $\mathcal{T}_{RL}$, lead to the same Bethe equations and the same eigenvectors but not the same eigenvalues. Nevertheless, both eigenvalues are related in a simple way $\Lambda_M^{(RL)}(\theta_k|\u|\vec{\theta}\, )= \left[ \Lambda_M^{(LL)}(\theta_k^*|\u^*|\vec{\theta}^*\, ) \right]^*$.

\section{\label{Recap} The composite model trick}

In this section we adapt the method described in \cite{PakuliakSlavnov} and \cite{JuanMiguel} to the present case. As the expressions appearing during the rest of this article usually involve several products, we will use the following shorthand notations when need to multiply over sets of rapidities
\begin{equation}
	h(\u )=\prod_{u_i\in \{\u\}} h(u_i) \ , \quad f(\u, \v)=\prod_{u_i\in \{\u\}} \prod_{v_j\in \{\v\}} f(u_i,v_j) \ , \quad  f^<(\u, \u )=\prod_{\substack{u_i,u_j\in \{\u\} \\ i<j}} f(u_i, u_j) \ .
\end{equation}
We will sometimes drop the arrows over the sets when the meaning is clear enough.

For the pure R-R flux model, the splitting of the scalar product into two smaller scalar products is given by
\begin{eqnarray}
&&	S_L( \u | \v )\equiv \langle 0 | C(\u ) B(\v ) | 0 \rangle = \nonumber\\
&&	\qquad \qquad \qquad \sum_{\substack{\{ \gamma \} \cup \{ \bar{\gamma} \}=\{ \u \} \\ \{ \beta \} \cup \{ \bar{\beta} \}=\{ \v \}}} (-1)^{|\beta | |\bar{\gamma} |} a_{l_1} (\bar{\beta} ) d_{l_1} (\bar{\gamma}) a_{l_2} (\gamma) d_{l_2} (\beta)   \frac{ S_{l_1} ( \gamma  | \beta ) S_{l_2} ( \bar{\gamma}  |  \bar{\beta} )}{\tanh \left( \frac{\gamma  - \bar{\gamma}}{2} \right)  \tanh \left( \frac{\bar{\beta } - \beta}{2} \right) } \ ,\label{breaking}
\end{eqnarray}
where $|\gamma|$ denotes the cardinality of the set and $l_1+l_2=L$ denote the lengths of the original spin chain and the two subchains. We should remark that the scalar products on the right-hand side do not vanish only when $|\beta|=|\gamma|$ and $|\bar{\beta}|=|\bar{\gamma}|$, so not all the possible partitions contribute and only scalar products with $|\u|=|\v|$ can be non-zero.

To prove this statement we break the monodromy matrix into two contributions in the following way\footnote{A similar statement holds for the opposite ordering of the $R$-matrix factors inside the monodromy matrix.}
\begin{align*}
\mathcal{T}^{(L)}_0 (u) &= R_{0,1} (u-\theta_1) R_{0,2} (u-\theta_2) \dots R_{0,L} (u-\theta_L) \\
&= \left[ R_{0,1} (u-\theta_1) \dots R_{0,l_1} (u-\theta_{l_1}) \right] \left[ R_{0,l_1 + 1} (u-\theta_{l_1 + 1}) \dots R_{0,L} (u-\theta_L) \right] \\
&= \mathcal{T}_0^{(l_1)} (u) \mathcal{T}_0^{(l_2)} (u) \ ,
\end{align*}
where the last line should be understood as product of matrices in the auxiliary space. Explicitly writing the Bethe operators we get
\begin{align}
	A_L (u) &= A_{l_1} (u) A_{l_2} (u) - B_{l_1} (u) C_{l_2} (u) \ , & B_L (u) &= A_{l_1} (u) B_{l_2} (u) + B_{l_1} (u) D_{l_2} (u) \ , \notag \\
	C_L (u) &= C_{l_1} (u) A_{l_2} (u)+ D_{l_1} (u) C_{l_2} (u) \ , & D_L (u) &= -C_{l_1} (u) B_{l_2} (u) + D_{l_1} (u) D_{l_2} (u) \ ,
\end{align}
where the minus signs come from the grading factor associated to moving the auxiliary space factor in $\mathcal{T}^{(l_2)}$ through the physical space factors of $\mathcal{T}^{(l_1)}$. These expressions provide us with a recipe to split one isolated operator when we cut a spin chain into two, but we need a recipe for the splitting of several operators. Let us show first how to split the product of two $B$ operators and later argue how the result should generalise for a product of three or more.

To compute the splitting of two $B$ operators we break each of the operators and apply the RTT commutation relations. We can represent the two factors by a graded tensor product of physical spaces instead of using the $l_i$ sub-indices to easily keep track of minus signs from grading
\begin{align*}
	B(u) B(v)&=\left[ A (u) \otimes B (u) + B (u) \otimes D (u) \right] \left[ A (v)\otimes B (v) + B (v) \otimes D (v) \right]\\
	&= A(u) A(v) \otimes B(u) B(v) - A(u) B(v) \otimes B(u) D(v) + B(u) A(v) \otimes D(u) B(v) \\
	&+ B(u) B(v) \otimes D(u) D(v)= A(u) A(v) \otimes B(u) B(v)+ B(u) B(v) \otimes D(u) D(v) \\
	&+\coth \left( \frac{u-v}{2} \right) B(v) A(u) \otimes B(u) D(v) + \coth \left( \frac{v-u}{2} \right)  B(u) A(v) \otimes B(v) D(u) \ .
\end{align*}
We stress that the minus sign coming from the graded tensor product is necessary to cancel the unwanted terms from the RTT relations (\ref{AB})
\begin{align*}
	A(x) B(y) &= -\coth \left( \frac{x-y}{2} \right) B(y) A(x) +\csch \left( \frac{x-y}{2} \right) B(x) A(y) \ , \\
	D(x) B(u) &= \coth \left( \frac{y-x}{2} \right) B(y) D(x) -\csch \left( \frac{y-x}{2} \right) B(x) D(y) \ .
\end{align*}

This result can be generalised by direct computation for a larger number of $B$ operators, but it is simpler to argue how the formula we have obtained should scale based on the structure of the $R$-matrix. Because $R$-matrices affect only two spaces and leave unaltered the rest, the weights should only be a product of weights involving two rapidities. The general behaviour when applied to a right pseudo-vacuum is then given by
\begin{displaymath}
	B(\v) |0 \rangle = \sum_{\{\beta\} \cup \{\bar{\beta}\}=\v} w_B( \beta , \bar{\beta}) B(\beta) \otimes B(\bar{\beta})|0 \rangle \qquad \text{where} \qquad w_B( \beta , \bar{\beta})=\coth \left( \frac{\bar{\beta}-\beta}{2} \right) \alpha_{l_1} (\bar{\beta}) \delta_{l_2} (\beta) \ .
\end{displaymath}

The computation for the case of $C$ operators applied to a left pseudo-vacuum is done similarly, giving us at the end $w_C( \beta , \bar{\beta})=\coth \left( \frac{\beta-\bar{\beta}}{2} \right) \delta_{l_1} (\bar{\beta}) \alpha_{l_2} (\beta)$.

Finally, in order to apply these breaking rules to the scalar product, we have to include an extra sign due to the graded tensor product
\begin{align}
\label{scalo}
	&S_L (\u , \v ) = \sum_{\text{part.}} w_C (\gamma, \bar{\gamma} ) w_B(\beta, \bar{\beta} )  \langle 0 | \left[ C( \gamma) \otimes C(\bar{\gamma}) \vphantom{B(\beta ) \otimes B( \bar{\beta} )} \right] \left[ B(\beta ) \otimes B( \bar{\beta} ) \right] | 0 \rangle \nonumber \\
	&= \sum_{\text{part.}} w_C (\gamma, \bar{\gamma} ) w_B (\beta, \bar{\beta} ) (-1)^{|\bar{\gamma} | |\beta|} \langle 0 | C( \gamma) B(\beta ) \otimes C( \bar{\gamma}) B( \bar{\beta} ) | 0 \rangle \ .
\end{align}
Here ``part" denotes all the partitions of rapidities of type $u$ and of type $v$ in two sets each such that $|\gamma|=|\beta|$ because, otherwise, the scalar products in the subchains will vanish.  Note that, because $|\bar{\gamma}|=M-|\gamma|$, $|\gamma|=|\beta|$ and $(-1)^{|\gamma |^2}=(-1)^{|\gamma |}$, the sign factor here and in equation (\ref{breaking}) can be written as $(-1)^{|\beta | (M-1)}$ and so it only contributes when the total number of excitations is even.

One can use equation (\ref{breaking}) in two different ways, one can either follow \cite{PakuliakSlavnov} to compute off-shell scalar products by expanding the scalar products into different powers of $\alpha$ and $\delta$ functions
\begin{equation}
	\langle 0 |\prod_{j=1}^M C(u_j) \prod_{j=1}^M B(v_j) |0\rangle=\sum_{\substack{\{ \gamma \} \cup \{ \bar{\gamma} \}=\{ u \} \\ \{ \beta \} \cup \{ \bar{\beta} \}=\{ v \}}} W_{part} ( \{\gamma \} , \{ \bar{\gamma} \} | \{\beta \} , \{ \bar{\beta} \} ) \delta(\gamma) \alpha (\bar{\gamma}) \alpha(\beta) \delta(\bar{\beta}) \ ,
\end{equation}
or one can follow \cite{JuanMiguel} to compute on-shell norms by expanding the norms into different powers of the length of the spin chain
\begin{equation}
	\prod_k \lim_{u_k\rightarrow v_k} \langle 0 |\prod_{j=1}^M C(u_j) \prod_{j=1}^M B(v_j) |0\rangle=\sum_{j=1}^M \sigma_l^{(M)}(\u ) L^j \ .
\end{equation}

Similar tricks can be applied to the mixed flux model. In fact, the computation in the case of mixed flux is structurally similar to the case of pure R-R flux, as the commutation relations we get from the RTT relations are the same except for the use of different coefficient functions. The only important difference arises from the fact that the commutation relations between $B$ operators (and $C$ operators) are not of purely bosonic nor fermionic type. To circumvent this problem we have to use the symmetric eigenstates defined in equation (\ref{symmetrizedstates}) and the properties $Y(\theta ) =c(\theta ) Y(-\theta )$ and $d(\theta )=-c(\theta ) d(-\theta)$, which make the quotient $\frac{Y(\theta)}{d(\theta)}$ a suitable object to keep track of the minus sings from the graded tensor product. Thus, after a similar computation, the final expressions for the weights are
\begin{align*}
	w_C (\gamma , \bar{\gamma}) &= \frac{Y(\bar{\gamma}, \gamma )}{d(\bar{\gamma} , \gamma )} \ , & w_B (\beta , \bar{\beta}) &= \frac{Y(\beta ,\bar{\beta})}{d(\beta , \bar{\beta} )} \ ,
\end{align*}
while the splitting formula for the scalar product can be written as
\begin{equation}
S_L( \u | \v )= \sum_{\substack{\{ \gamma \} \cup \{ \bar{\gamma} \}=\{ \u \} \\ \{ \beta \} \cup \{ \bar{\beta} \}=\{ \v \}}} (-1)^{|\beta | |\bar{\gamma} |} a_{l_1} (\bar{\beta} ) d_{l_1} (\bar{\gamma}) a_{l_2} (\gamma) d_{l_2} (\beta)   \frac{ Y (\bar{\gamma}- \gamma) Y (\beta- \bar{\beta})}{d (\bar{\gamma} - \gamma ) d(\beta - \bar{\beta}) } S_{l_1} ( \gamma  | \beta ) S_{l_1} ( \bar{\gamma}  |  \bar{\beta} ) \ , \label{breaking2} 
\end{equation}
where $S_L( \u | \v )\equiv\langle \u | \v \rangle$ is the scalar product of symmetrised state and a symmetrised dual state.

\section{\label{Conj} Conjugation of the Bethe operators}

We would like to compute the norm of the Bethe states which are constructed by means of the Algebraic Bethe Ansatz summarised in section \ref{ABA}. In order to do this, we need to calculate\footnote{We shall from now on often suppress explicitly indicating the inhomogeneities in the argument of the operators, when the context will be sufficiently clear.}
\begin{equation}
\langle 0|B^\dagger(u_1)...B^\dagger(u_M)B(u_M)...B(u_1)|0\rangle \ ,
\end{equation}
which gives the squared-norm of a state with $M$ magnon excitations when the magnon rapidities $u_i$ are all taken as real for all $i=1,...,M$. The first task is therefore to make sense of the Hermitian conjugate of the magnon creation-operator $B(u|\vec{\theta})$, which can be done following \cite{AnastasiaRafael}.

\subsection{Pure R-R case}

Let us start with the case of the $R$-matrix for the pure R-R case (\ref{matrR}). We first notice that such matrix satisfies
\begin{equation}
\label{st}
R^{st_0 st_i}_{0,i} (\theta) = R_{0,i}(\theta) \ , \qquad \rho_0^{-1} R^{st_0}(\theta+ i \pi) \rho_0 = \frac{1}{b(\theta)} R(\theta) \ , \qquad \rho = \begin{pmatrix}0&1\\i&0
\end{pmatrix} \ ,
\end{equation}
where the apex $st_i$ denotes supertransposition in the $i$-th space\footnote{We work in the conventions $E_{ij}^{st} = (-)^{[i][j] + [i]} E_{ji}$.} and $\rho_0^{\pm 1}$ denotes $\rho^{\pm 1} \otimes \mathfrak{1}$, where $\mathfrak{1}$ is the two-by-two identity matrix. The $R$-matrix also satisfies $R^2(\theta)=\mathfrak{1}\otimes \mathfrak{1}$ and $R^* (\theta)=R (\theta^*)$.

With this information, we proceed as follows: We first compute the supertranspose of the monodromy matrix in all the quantum spaces
\begin{equation}
{\cal{T}}_L ^{st_{\text{phys}}}(u) = \prod_{i=1}^L R_{0,i}^{st_i}(u - \theta_i) \ ,
\end{equation}
we then use (\ref{st}) to trade this for a supertransposition in the auxiliary space (remembering that the supertransposition is not an involution)
\begin{equation}
{\cal{T}}_L ^{st_{\text{phys}}}(u) = \prod_{i=1}^L F_0 R_{0,i}^{st_0}(u - \theta_i) F_0 \ ,
\end{equation}
where $F=$diag$(1,-1)$.
Using (\ref{st}) again, it is easy to see that this can be simplified as follows
\begin{equation}
\label{readoff}
{\cal{T}}_L ^{st_{\text{phys}}}(u) = F_0 \rho_0 \Big[\prod_{i=1}^L \frac{1}{b(u - \theta_i - i \pi)} \, R_{0,i}(u - \theta_i - i \pi)\Big] \rho_0^{-1} F_0= \Big[\prod_{i=1}^L \frac{1}{b(u - \theta_i - i \pi)}\Big] \rho_0^* {\cal{T}}(u - i \pi) (\rho_0^*)^{-1} \ ,
\end{equation}
where we have telescopically cancelled the matrices $F_0 \rho_0$ and their inverses appearing near one another in the intermediate terms, and have taken into account the fermionic signs generated by tensor-product multiplication. Given the form of the matrix $\rho$ in (\ref{st}), it is now possible to read off from (\ref{readoff}) the relationship we are looking for, properly taking into account the fermionic nature of $\rho$ and the $C$ operator\footnote{The function $b(\theta)$ has zeros on the imaginary axis for special values of the argument (and at infinity). We shall be considering  regions away from such special points of the arguments appearing in (\ref{complex}).}
\begin{equation}
\label{complex}
B^{st_{\text{phys}}}(u) = - i \, C(u - i \pi) \, \prod_{i=1}^L \frac{1}{b(u - \theta_i - i \pi)} \ .
\end{equation}
Finally, we need to implement complex conjugation to define an appropriate hermitian conjugate:
\begin{equation}
\label{defin}
B^\dagger(u) \equiv  - [B^{st_{\text{phys}}}(u)]^* \ .
\end{equation} 
It is not unexpected to have to rectify a minus sign to ensure positive norms, given the sign one accumulates naturally in doing the supertransposition; the same will happen for the mixed-flux case later on.

By taking the complex conjugate of (\ref{complex}) and using that $B(u)^*=B(u^*)$, we therefore obtain\footnote{Instead of taking the complex conjugation at the end of the computation we could have done it at the beginning of it, getting instead $i \, C(u^* - i \pi| \{\theta_j^*\}) \, \prod_{i=1}^L b^{-1} (u^* - \theta_i^* - i \pi)$. Both expressions are equal when one takes into account the explicit expression of $b(\theta)$ and the quasi-periodicity of the $C$ operator, $C(\theta\pm 2 \pi i)=-C(\theta)$.}
\begin{equation}
\label{multi}
B^\dagger(u| \{\theta_j\}) = \frac{- i \, C(u^* + i \pi | \{\theta_j^*\}) }{ \prod_{i=1}^L b(u^* - \theta_i^* + i \pi)}= \frac{ - i \, C(u^* | \{\theta_j^*- i \pi\})}{\prod_{i=1}^L b(u^* - \theta_i^* + i \pi)} \ .
\end{equation}

From this relation we can distinguish between two particular cases of relevance amongst the many possible, the case where Im$[\theta_j]=0$ for all $j$ and the case where Im$[\theta_j]=-i \pi/2$ for all $j$. Assuming that $u$ is real we will have $B^\dagger (u) \propto C(u + i \pi)$ in the first (or {\it asymmetric}) case while $B^\dagger (u) \propto C(u)$ in the second (or {\it symmetric}) one. The second case is more akin to an $SU(1|1)$ version of the usual XXZ spin chain, which allows us to borrow some intuition from it. Some explicit computations of norm in these two cases are collected in appendix~\ref{new}.

As a final remark about this case, (\ref{readoff}) also imply $A^\dagger(u) |0\rangle = A(u)|0\rangle$ and $D^\dagger(u) |0\rangle = D(u)|0\rangle$, which is consistent with the consideration that $\langle 0|A(u)|0\rangle$ and $\langle 0|D(u)|0\rangle$ should unambiguously be given by the respective vacuum-eigenvalues.

\subsection{Mixed flux case}

As we commented before, to obtain the Hermitian conjugate of the $B$ operator in the case of mixed flux we need to resort to introduce the $R$ representation and the monodromy matrix $\mathcal{T}_{RL}$. In the pure R-R case we did not need to resort to this representation thanks to the accidental relation (\ref{st}), which allowed it to remain in the same representation after supertransposing one of the spaces, but this is no longer possible with the $LL$ $S$-matrix because $R_{LL}^{st_1}(\theta)$ is proportional to $R_{RL}(\theta)$.

The $RL$ $S$-matrix, for the scattering of a $R$ with a $L$ representation - in a suitable normalisation, ignoring the dressing factor which is inessential to our discussion - is given by\footnote{The functions involved in $R_{RL}$ can be written in terms of the ones involved in $R_{LL}$, e.g. $a_{RL}(\theta)=Y(\theta)$, but we prefer to denote them independently for the sake of clarity.}
\begin{equation}
R_{RL}(\theta) = a_{RL} (\theta) E_{11} \otimes E_{11} + E_{11} \otimes E_{22} + b_{RL} (\theta) \left[ E_{21} \otimes E_{21} - E_{12} \otimes E_{12} \right] + E_{22} \otimes E_{11} + c_{RL} (\theta) E_{22} \otimes E_{22} \ , \label{RmatrixRL}
\end{equation}
\begin{equation}
\label{below}
a_{RL} (\theta) = - i \frac{\sin\Big(\frac{\pi}{k} + \frac{i \theta}{2}\Big)}{\sinh \frac{\theta}{2}} \ , \qquad b_{RL} (\theta) = \frac{\sin \frac{\pi}{k}}{\sinh \frac{\theta}{2}} \ , \qquad c_{RL} (\theta) = \cos \frac{\pi}{k} + i \coth \frac{\theta}{2} \, \sin \frac{\pi}{k},
\end{equation}
where we have introduced a momentum $p$ associated to the (negative) momentum $p_1<0$ of the $R$ representation to be $p = - p_1 >0$ (see the discussion in \cite{Fontanella:2019ury}) and we have set $-p = e^{\theta_1}$. The momentum $p_2>0$ of the $L$ representation is then defined as $p_2 = e^{\theta_2}$. Finally, $\theta = \theta_1 - \theta_2$.

This $R$-matrix fulfils a mixed Yang-Baxter equation with the $R$-matrix for the $LL$ representation, which can be used to compute the commutation relations between the monodromy matrices $\mathcal{T}_{RL}$ and $\mathcal{T}_{LL}$
\begin{equation}
R_{RL}(\theta - \theta') \, [\mathcal{T}_{RL}]_1(\theta) \, [\mathcal{T}_{LL}]_2(\theta') = [\mathcal{T}_{LL}]_2(\theta') \, [\mathcal{T}_{RL}]_1(\theta) \, R_{RL}(\theta - \theta'),  
\end{equation}
from which the particular relations we require in order to compute scalar products and norms reads as follows\footnote{The last expression is not obtained directly from an individual RTT relation but combining two of them. In particular $\tilde {D} (\theta) B(\theta ')= c_{RL} (\theta - \theta') B(\theta ') \tilde {D} (\theta) -b_{RL} (\theta - \theta') A(\theta ') \tilde {C} (\theta)$ and $\tilde {C} (\theta) A(\theta ')= a_{RL} (\theta - \theta') A(\theta ') \tilde {C} (\theta) -b_{RL} (\theta - \theta') B(\theta ') \tilde {D} (\theta)$, together with the relation $a_{RL} c_{RL} - b_{RL}^2=1$.}
\begin{align}
\tilde{B}(\theta) B(\theta ') &= \frac{b_{RL} (\theta - \theta')}{a_{RL} (\theta - \theta')} \big[A(\theta ') \tilde{A}(\theta) - \tilde{D}(\theta) D(\theta ')\big] - \frac{c_{RL} (\theta - \theta')}{a_{RL} (\theta - \theta')} B(\theta ') \tilde{B}(\theta) \ , \notag  \\
\tilde {A} (\theta) B(\theta ') &= \frac{1}{a_{RL} (\theta - \theta')} B(\theta ') \tilde {A} (\theta) -\frac{b_{RL} (\theta - \theta')}{a_{RL} (\theta - \theta')} \tilde {C} (\theta)  D(\theta ')\ ,  \notag \\
\tilde {D} (\theta) B(\theta ') &= \frac{1}{a_{RL} (\theta - \theta')} B(\theta ') \tilde {D} (\theta) -\frac{b_{RL} (\theta - \theta')}{a_{RL} (\theta - \theta')} \tilde {C} (\theta)  A(\theta ') \ . \label{commutationRLwithLL}
\end{align}

We can now report the crossing equation that was used in \cite{Fontanella:2019ury} to derive constraints on the dressing phases associated to the $S$-matrices for the various combinations of representations
\begin{equation}
\label{st2}
R_{LL}(\theta) \, \sigma_0^{-1} R_{RL}^{st_0}(\theta) \, \sigma_0 = a_{RL}(\theta) \mathfrak{1} \otimes \mathfrak{1} \ , \qquad  \sigma_0= \sigma \otimes \mathfrak{1} \ , \qquad \sigma = \begin{pmatrix}1&0\\0&-i
\end{pmatrix} \ ,
\end{equation}
where $R_{LL}$ is the mixed-flux $LL$ $S$-matrix previously introduced. We also have once again $R_{LL}^{st_0, st_i}(\theta) = R_{LL}(\theta)$ while $R_{RL}^{st_0}(\theta) = R_{RL}^{st_i}(\theta)$. 

We can now make use of a particular property of $R_{LL}$
\begin{equation}
R_{LL}(\theta) R_{LL}(-\theta) = \mathfrak{1} \otimes \mathfrak{1} \ ,
\end{equation}
from which we get
\begin{equation}
\label{st22}
\sigma_0^{-1} R_{RL}^{st_0}(\theta) \, \sigma_0 = a_{RL} (\theta) R_{LL}(-\theta) \ .
\end{equation}

With this information, we once again proceed as in \cite{AnastasiaRafael}. We first compute the supertranspose of the monodromy matrix in all the quantum spaces. For the moment we will use the standard definition monodromy matrix instead of the rescaled version presented in equation (\ref{transfer-monodromy-RL})
\begin{equation}
\mathbb{T}_{RL}^{st_{\text{phys}}}(u|\vec{\theta}) = \prod_{i=1}^L R_{RL;0,i}^{st_i}(u - \theta_i) \ . 
\end{equation}
We then trade the supertransposition in physical space for a supertransposition in the auxiliary space for each $R$-matrix, with no extra matrix-similarity this time
\begin{equation}
\mathbb{T}_{RL}^{st_{\text{phys}}}(u|\vec{\theta}) = \prod_{i=1}^L R_{RL;0,i}^{st_0}(u - \theta_i) \ .
\end{equation}
It is easy to see that this can be simplified as follows
\begin{equation}
\label{readoff3}
\mathbb{T}_{RL}^{st_{\text{phys}}}(u|\vec{\theta}) = \sigma_0 \Bigg[ R_{LL;0,i}(\theta_i - u) \, \prod_{i=1}^L a_{RL}(u - \theta_i) \Bigg] \sigma_0^{-1} = \Bigg[\prod_{i=1}^L a_{RL} (u - \theta_i) \Bigg] \sigma_0 \mathcal{T}_{LL}\Big(-u| -\vec{\theta} \,\Big) \sigma_0^{-1} \ ,
\end{equation}
where we have telescopically cancelled the matrices $\sigma_0$ and their inverses appearing near one another in the intermediate terms. From this we get
\begin{equation}
\mathbb{B}^{st_{\text{phys}}}\Big(u| \vec{\theta} \,\Big) = i \, \Bigg[\prod_{i=1}^L a_{RL} (u - \theta_i) \Bigg] \, B\Big(-u| -\vec{\theta} \,\Big) \ . 
\end{equation}

Instead of carrying the factor $\prod a_{RL} (u - \theta_i)$ though all the computation, it is simpler if we redefine the monodromy matrix by including this factor inside as we can pair each factor of $a_{RL}$ with an $R$-matrix factor
\begin{equation}
	\mathcal{T}_{RL} (u|\vec{\theta})=\mathbb{T}_{RL} (u|\vec{\theta}) \prod_{i=1}^L \frac{1}{a_{RL} (u - \theta_i)}=\prod_{i=1}^L \frac{R_{RL;0,i}(\beta - \theta_i)}{a_{RL} (u - \theta_i)} \ .
\end{equation}
It is immediate to repeat the same steps as before with this new definition of the monodromy matrix, giving us at the end
\begin{equation}
\label{consti}
\tilde{B}^{st_{\text{phys}}}\Big(u| \vec{\theta} \,\Big) = i \, B\Big(-u| -\vec{\theta} \,\Big) \ . 
\end{equation}

On the other hand, we could have started from this other equation:
\begin{equation}
\label{st3}
\sigma_0^{-1} R_{RL}(\theta) \, \sigma_0 = a_{RL} (\theta) R_{LL}^{st_1}(\theta) \ . 
\end{equation}
From here, a procedure completely analogous to the one just performed leads to
\begin{equation}
\label{readoff2}
\mathcal{T}_{LL}^{st_{\text{phys}}}(u|\vec{\theta}) = F_0 \, \sigma_0 \mathcal{T}_{RL}\Big(-u| -\vec{\theta} \,\Big) \sigma_0^{-1} \, F_0 \ ,
\end{equation}
where we have used the property $F_0 R_{LL}^{st_0}(\theta) F_0 = R_{LL}^{st_i}(\theta)$. Therefore, we obtain
\begin{equation}
B^{st_{\text{phys}}}\Big(u| \vec{\theta} \,\Big) = i \, \tilde{B}\Big(-u| -\vec{\theta} \,\Big) \ , 
\end{equation}
which is consistent with (\ref{consti}) if we take into account the non-involutivity of the supertranspose, which calls into the game the matrix $F$ and adds an extra sign to this equation.

Finally, these equations imply
\begin{equation}
B^\dagger (u)= -\left[ B^{st_{\text{phys}}}\Big(u | \vec{\theta} \,\Big) \right]^*= i  \, \left[ \tilde{B}\Big(-u | -\vec{\theta} \,\Big) \right]^* \ , \label{mixedfluxconjugation}
\end{equation}
where the extra minus sign has been justified in the pure R-R discussion.
The operator $\tilde{B}$ is now inherently complex, since it is built out of the complex coefficient functions appearing in the mixed flux $R$-matrix. By inspection, we see that $a_{RL}$ and $c_{RL}$ satisfy
\begin{equation}
\left[ a_{RL}(\theta) \right]^* = a_{RL}(-\theta^*) \ , \qquad \left[ c_{RL}(\theta) \right]^* = c_{RL}(-\theta^*) \ ,
\end{equation}
while
\begin{equation}
\left[ b_{RL}(\theta) \right]^* = - b_{RL}(-\theta^*) \ .
\end{equation}
Nevertheless, the only terms in the $R$-matrix which allow creation and annihilation of magnons come equipped with a $b_{RL}$, hence the $\tilde{B}$ will always have an odd number of $b_{RL}$ in each of its terms. \footnote{This is because, when writing the $B$ operator in terms of rising and lowering Pauli matrices, it takes the schematic form $\sum_j s_j \sigma_+^{j+1} \sigma_-^j$, so we can expect the factors $s_j$ to have $2j+1$ factors of $b_{RL}$. In the sum all the rising and lowering matrices act on different sites and the sum is over all the possible ways they can act.} This means that
\begin{equation}
B^\dagger (u)= i \left[ \tilde{B}\Big(-u| -\vec{\theta} \,\Big) \right]^*= -i \, \tilde{B}\Big(u^*| \vec{\theta}^* \,\Big) \ . \label{mixedfluxconjugation2}
\end{equation}

\section{\label{Offshell}Off-shell scalar product in the pure R-R case}

In this section we compute the scalar product between two off-shell Bethe vectors for the $R$-matrix given in equation (\ref{matrR}). Here and in the next section we will follow closely the method explained in \cite{PakuliakSlavnov} to accomplish our task. 

In order to construct the scalar product we first have to classify the terms which appear in it by singling out the factors of $\alpha$ and $\delta$
\begin{equation}
	\langle 0 |\prod_{j=1}^M C(u_j) \prod_{j=1}^M B(v_j) |0\rangle=\sum_{\substack{\{ \gamma \} \cup \{ \bar{\gamma} \}=\{ \u \} \\ \{ \beta \} \cup \{ \bar{\beta} \}=\{ \v \}}} W_{part} ( \{\gamma \} , \{ \bar{\gamma} \} | \{\beta \} , \{ \bar{\beta} \} ) \delta (\gamma) \alpha (\bar{\gamma}) \alpha(\beta) \delta(\bar{\beta}) \label{generalform} \ .
\end{equation}

The coefficients $W_{part}$ are usually very difficult to compute, except for the two highest coefficients, that is,
\begin{equation}
	Z(\u |\v ) =W_{part} ( \emptyset , \u | \emptyset , \v ) \ , \qquad \text{ and } \qquad \bar{Z}(\u |\v ) =W_{part} ( \u  , \emptyset  | \v , \emptyset ) \ .
\end{equation}
In particular, using the commutation relations and symmetry properties, we obtain
\begin{align}
	Z (\u |\v ) &= (-1)^{\frac{|u|(|u|-1)}{2}} \csch \left( \frac{ \u - \v}{2} \right) \cosh^< \left( \frac{ \u - \u}{2} \right) \cosh^< \left( \frac{ \v - \v}{2} \right) \ , \notag \\
	\bar{Z} (\u | \v ) &= (-1)^{\frac{|u|(|u|-1)}{2}} \csch \left( \frac{ \v - \u}{2} \right) \cosh^< \left( \frac{ \u - \u}{2} \right) \cosh^< \left( \frac{ \v - \v}{2} \right) \label{HighestweightRR} \ .
\end{align}
The details of this computation are relegated to appendix \ref{highestweights1}.

The remaining coefficients can be obtained from the explicit expression of the highest coefficients and the composite model splitting of the scalar product (\ref{breaking}). The procedure to construct these coefficients is based on the fact that they only depend on the form of the $R$-matrix but not on the particular form of the $\alpha (u)$ and $\delta (u)$ functions - which can be thought of as corresponding to different representations of the RTT algebra. This independence from $\alpha$ and $\delta$ can be proven by examining the structure of the scalar product under application of the commutation relations, where $W_{part}$ will just collect the functions arising from the commutation relations and no operator $A$ and $D$. Thus, we can compute the scalar product for a different and non-physical model where the calculations are particularly simple, and extract the coefficients from this model to use them in the one of our interest. By a stroke of luck, we can choose the functions $\alpha$ and $\delta$ in such a way that only one of the terms in (\ref{breaking}) and only one of the terms in (\ref{generalform}) contribute.

If we choose a model such that $\alpha_{l_1} (u_i)=0$ if $u_i\in \hat{\gamma}$ and $\alpha_{l_2} (v_i)=0$ if $v_i \in \hat{\beta}$, such that $|\hat{\gamma}|+|\hat{\beta}|=M$, then just one term contributes to equation (\ref{breaking}). The only term that can appear is the one with $\gamma=\hat{\gamma}$ and $\bar{\beta}=\hat{\beta}$. If we choose $\gamma$ such that it contains further rapidities apart from those in $\hat{\gamma}$, then $\beta$ should contain some rapidities from $\hat{\beta}$, so such contribution vanishes. This happens because we cannot fit all rapidities from $\hat{\beta}$ in $\bar{\beta}$, as $|\bar{\beta}|=M-|\beta |=M-|\gamma|<M-|\hat{\gamma}|=|\hat{\beta}|$. Other possibilities can be ruled out after a similar reasoning. In addition, if we use that $\delta_L (x)=\delta_{l_1} (x) \delta_{l_2} (x)$ we can see that only one term from equation (\ref{generalform}) can contribute, by using a similar reasoning. Notice also that the smaller scalar products we get in equation (\ref{breaking}) will only contribute with the highest and lowest weight coefficients.

Equating the only term we get from equation (\ref{generalform}) and the only term we get from equation (\ref{breaking}) and cancelling the factors of $\alpha$ and $\delta$, we obtain the expression
\begin{equation}
	W_{part} (\gamma , \bar{\gamma} | \beta , \bar{\beta} )  = (-1)^{|\gamma| (M-|\gamma|)} \frac{\bar{Z}(\gamma | \beta) Z (\bar{\gamma} | \bar{\beta} )}{\tanh \left( \frac{\bar{\gamma}  - \gamma}{2} \right)  \tanh \left( \frac{\beta - \bar{\beta } }{2} \right) } \ .
\end{equation}
Substituting here the explicit expressions for the highest and lowest weight from \ref{HighestweightRR}
\begin{align}
	W_{part} (\gamma , \bar{\gamma} | \beta , \bar{\beta} ) & = (-1)^{\frac{|\gamma| (|\gamma|-1)}{2}+\frac{(M-|\gamma|) (M-|\gamma|-1)}{2}+|\gamma| (M-|\gamma|)} \cosh^< \left( \frac{ \u - \u}{2} \right) \cosh^< \left( \frac{ \v - \v}{2} \right) \notag \\
	&\times  \csch \left( \frac{  \gamma - \bar{\gamma} }{2} \right) \csch \left( \frac{ \bar{\beta} - \beta }{2} \right) \csch \left( \frac{ \beta - \gamma }{2} \right) \csch \left( \frac{ \bar{\gamma} - \bar{\beta}}{2} \right)  \ .
\end{align}
It is easy to see that the sign only contributes at the end with a global factor $(-1)^{\frac{M (M-1)}{2}}$.

We can now substitute back into (\ref{generalform}) and perform the sum over partitions of these weights with $\delta (\gamma) \delta (\bar{\beta})$. First, we can take the factors $\cosh^< \left( \frac{ \u - \u}{2} \right) \cosh^< \left( \frac{ \v - \v}{2} \right)$ out of the sum over partitions as they are invariant under a permutation of the rapidities. This allows us to apply a corollary to the hyperbolic version of the Cauchy determinant, equation~(\ref{corollary}), and cast the scalar product as a determinant
\begin{equation}
\langle 0 |\prod_{j=1}^M C(u_j) \prod_{j=1}^M B(v_j) |0\rangle=\coth^< \left( \frac{ \v - \v }{2} \right) \coth^< \left( \frac{ \u - \u }{2} \right) \det \left[ \csch \left( \frac{ u_i - v_j }{2} \right) \left[ \delta (v_j) - \delta (u_i)\right] \right] \ , \label{offshellscalarproductnoflux}
\end{equation}
where we have used that $(-1)^{\frac{M (M-1)}{2}} \coth^> \left( \frac{ \v - \v }{2} \right)=\coth^< \left( \frac{ \v - \v }{2} \right)$.

\section{Off-shell scalar product in the mixed flux case}

In this section we compute the scalar product of two states for the case of mixed flux. In this case we can either use operators built with the $\mathcal{T}_{LL}$ monodromy matrix or with the $\mathcal{T}_{RL}$ monodromy matrix. We will focus here on the case where all operators are built using the $\mathcal{T}_{LL}$ monodromy matrix, and the case where all creation operators are built using $\mathcal{T}_{LL}$ and all annihilation operators are build using $\mathcal{T}_{RL}$. The case where all operators are built using the $\mathcal{T}_{RL}$, and the case where all creation operators are built using $\mathcal{T}_{RL}$ and all annihilation operators are built using $\mathcal{T}_{LL}$, can respectively be obtained from these two via the relation (\ref{readoff3}). We will not treat here the most general case involving all four kinds of creation and annihilation operators at the same time.

\subsection{Scalar product of LL-LL Bethe operators}

In this subsection we will focus in the case where both states are constructed using the $\mathcal{T}_{LL}$ monodromy matrix. Although this is not the correct scalar product we need to compute in order to construct the norm of a state, it is useful to know it nonetheless.

Because the steps and computations involved here are very similar to those in the previous section, we only write the final expressions for each step while highlighting the differences between the two computations.

First, we have to clarify that in this subsection we will define the scalar product using the symmetrised version of the states
\begin{align}
	\langle \u | \v \rangle &= \frac{\langle 0 | C(u_1|\vec{\theta}\, ) \dots C(u_M|\vec{\theta}\, ) B(v_M|\vec{\theta}\, ) \dots B(v_1 |\vec{\theta}\, )|0\rangle}{d^> (\u - \u) d^> (\v - \v)} \label{mixedscalarproductdefinition} \\
	&= \sum_{\substack{\{ \gamma \} \cup \{ \bar{\gamma} \}=\{ \u \} \\ \{ \beta \} \cup \{ \bar{\beta} \}=\{ \v \}}} W_{part} ( \{\gamma \} , \{ \bar{\gamma} \} | \{\beta \} , \{ \bar{\beta} \} ) \delta (\gamma) \delta(\bar{\beta}) \ . \notag
\end{align}

The following step is to compute the highest weight contribution to the scalar product \emph{without} including the symmetrisation factors $d^> (\u - \u) d^> (\v - \v)$. We will relegate again this computation to appendix \ref{highestweights2} and reproduce here the final result, equation (\ref{highestweightmixed}),
\begin{equation}
	Z_M (\u | \v)= (-1)^M \bar{Z}_M (\u | \v) = (-1)^{\frac{|u|(|u|-1)}{2}} \frac{Y^> (\u - \u) Y^> (\v - \v )}{X^> (\u - \u) X^> (\v - \v )} X(\u - \v ) \ .
\end{equation}

With this function and the recipe to split the spin chain into two showed in equation (\ref{breaking2}),  we can write the general weight for the scalar product
\begin{align*}
	&W(\gamma ; \bar{\gamma} | \beta ; \bar{\beta} ) =(-1)^{|\gamma | |\bar{\beta}|} w_B (\bar{\beta};\beta) w_C (\bar{\gamma}; \gamma ) \bar{Z} (\beta | \gamma) Z (\bar{\gamma} | \bar{\beta} )= (-1)^{\frac{M (M-1)}{2}} \frac{ Y (\bar{\beta} - \beta) Y (\gamma - \bar{\gamma})}{d (\bar{\beta} - \beta) d(\gamma - \bar{\gamma})} \times \\
	&  \frac{Y^> (\gamma - \gamma) Y^> (\beta - \beta )}{X^> (\gamma - \gamma) X^> (\beta - \beta )} \frac{X(\beta - \gamma)}{d^> (\gamma - \gamma) d^> (\beta - \beta)} \frac{Y^> (\bar{\gamma} - \bar{\gamma}) Y^> (\bar{\beta} - \bar{\beta} )}{X^> (\bar{\gamma} - \bar{\gamma}) X^> (\bar{\beta} - \bar{\beta} )} \frac{X(\bar{\gamma} - \bar{\beta} )}{d^> (\bar{\gamma} - \bar{\gamma}) d^> (\bar{\beta} - \bar{\beta} )} \ .
\end{align*}
Notice that we have divided the highest weights by factors of $d$ to adapt them to the correct form of the scalar product we are computing. After reordering of some arguments using that $Y(\theta)/d(\theta)=-Y(-\theta)/d(-\theta)$ and taking the factors $\frac{Y^>}{X^> d^>}$ out of the sum over partitions, which we are allowed to do because this combination is invariant under permutation of the rapidities, we get
\begin{multline*}
	S(\u | \v)=(-1)^{\frac{M (M-1)}{2}} \frac{Y^> (\u - \u) Y^> (\v - \v)}{d^>( \u - \u) d^>(\v - \v) X^> (\u - \u ) X^> (\v - \v )} \\
	\sum_{\text{part.}} X(\gamma -\beta) X(\bar{\gamma} - \bar{\beta} ) X( \bar{\beta} -\beta ) X(\gamma - \bar{\gamma} ) d(\gamma) d(\bar{\beta} )\ .
\end{multline*}
At this point the sum over partitions has the correct form to apply equation (\ref{corollary}). This allows us to cast the scalar product as a determinant
\begin{equation}
	S(\u | \v)=\frac{Y^> (\u - \u) Y^> (\v - \v)}{d^>( \u - \u) d^>(\v - \v)}  \det \left[ \vphantom{\frac{a}{b}} X(u-v) [\delta (v) - \delta (u)] \right] \ .  \label{offshellscalarproductmixedflux}
\end{equation}
At this point we can check that this scalar product has the correct symmetry properties under the exchange of rapidities inherited from the original states. Furthermore, we can see that the $d$ functions appear in the same form as in the definition (\ref{mixedscalarproductdefinition}), thus we can get rid of them and write
\begin{equation}
	\langle 0 | C(u_1|\vec{\theta}\, ) \dots C(u_M|\vec{\theta}\, ) B(v_M|\vec{\theta}\, ) \dots B(v_1 |\vec{\theta}\, )|0\rangle=Y^> (\u - \u) Y^> (\v - \v)  \det \left[ \vphantom{\frac{a}{b}} X(u-v) [\delta (v) - \delta (u)] \right] \ .
\end{equation}

\subsection{Scalar product of RL-LL Bethe operators}

In this section we will compute the scalar product between $\tilde{B}$ operators from the $RL$ monodromy matrix and $B$ operators from the $LL$ monodromy matrix. This is the correct scalar product we have to compute if we want to obtain the norm of the states, as we have previously derived that $B^\dagger \propto \tilde{B}$. 

First, we should clarify what quantity we will compute. In this case we will divide the scalar product of $\tilde{B}$ and $B$ operators by a factor  $d^< (\u - \u) d^> (\v - \v)$ to symmetrise the states and simplify our computations. In contrast with the previous section, where the two sets of $d$ functions were ordered in the same way in order to get a scalar product symmetric under the permutation of rapidities, here the two sets of $d$ functions have to be order oppositely because the $\tilde{B}$ operators satisfy
\begin{eqnarray}
&&\tilde{B}(u_1|\vec{\theta}\, ) \tilde{B}(u_2|\vec{\theta}\, ) +c_{LL} (u_1-u_2) \tilde{B}(u_2|\vec{\theta}\, ) \tilde{B}(u_1|\vec{\theta}\, ) =0 \ ,
\end{eqnarray}
which is exactly the same type of relations satisfied by the $B$ operators, however the chain of $B$'s and of $\tilde{B}$'s appear in the formula below in opposite order 
\begin{equation}
S( \u | \v)=\frac{\langle 0 | \tilde{B}(u_1|\vec{\theta}\, ) \dots \tilde{B}(u_M|\vec{\theta}\, ) B(v_M|\vec{\theta}\, ) \dots B(v_1 |\vec{\theta}\, )|0\rangle}{d^< (\u - \u) d^> (\v - \v)} \ .
\end{equation}
In addition, we also have to modify the ansatz for the scalar product as the commutation relation between $\tilde{B}$ and $B$ operators, (\ref{commutationRLwithLL}), is not exactly the same as the one between $C$ and $B$ operators. Nevertheless, the structure is similar enough that we are allowed to apply the same procedure as the previous two cases just by treating $\tilde{A}$ as a $D$ operator and $\tilde{D}$ as an $A$ operator
\begin{equation}
S( \u | \v)=\sum_{\substack{\{ \gamma \} \cup \{ \bar{\gamma} \}=\{ \u \} \\ \{ \beta \} \cup \{ \bar{\beta} \}=\{ \v \}}} W_{part} ( \{\gamma \} , \{ \bar{\gamma} \} | \{\beta \} , \{ \bar{\beta} \} ) \Delta(\bar{\gamma}) \delta(\bar{\beta}) \ .
\end{equation}
Here $\Delta$ depends on the set $\bar{\gamma}$ in contrast with the previous cases, where the function accompanying $W_{part}$ had a dependence on $\gamma$ instead. We can recover the usual dependence if we further divide $S( \u | \v)$ by $\Delta (\u)$.

From the commutation relations (\ref{commutationRLwithLL}) we can compute the highest and lowest weights for this setting as explained in appendix \ref{highestweights3}, with the final result being
\begin{equation}
	\hat{Z}_M = (-1)^M \hat{\bar{Z}}_M= (-1)^M \frac{ b_{RL} (\u - \v) a_{RL}^< (\u - \u ) a_{RL}^> (\v - \v )}{ a_{RL} (\u - \v) b_{RL}^< (\u - \u ) b_{RL}^> (\v - \v )} \ .
\end{equation}

We do not need to compute any splitting factors in this case as we already know them for the $B$ operators and the ones for the $\tilde{B}$ operators are the same as those, as their commutation relations are equivalent, 
\begin{equation}
	\frac{S( \u | \v)}{\Delta (\u)}= \sum_{\text{part.}} (-1)^{|\gamma| (M-|\gamma|)} \frac{Y(\bar{\gamma} - \gamma) Y(\bar{\beta} - \beta )}{d(\bar{\gamma} - \gamma ) d(\bar{\beta} - \beta )} \frac{\hat{\bar{Z}}(\gamma | \beta) \hat{Z}(\bar{\gamma} | \bar{\beta}) \Delta^{-1}(\gamma) \delta(\bar{\beta})}{d^< (\gamma - \gamma) d^< (\bar{\gamma} - \bar{\gamma}) d^> (\beta - \beta) d^> (\bar{\beta} - \bar{\beta}) } \ .
\end{equation}
From here we can exchange the order of the $\frac{Y(u_{II} , u_{I})}{d(u_{II} , u_I )}$ to get rid of the $(-1)^{|u_I| (M-|u_I|)}$ sign and take the factors $\frac{Y}{d b_{RL}}$ out of the sum over partitions
\begin{displaymath}
	\frac{S( \u | \v)}{\Delta (\u)}=\frac{ Y^< (\u - \u ) Y^> (\v - \v )}{d^< (\u - \u ) d^> (\v - \v ) b_{RL}^< (\u - \u ) b_{RL}^> (\v - \v )} \sum_{\text{part.}} \frac{b_{RL} (\gamma , \beta) b_{RL} (\bar{\beta} , \bar{\gamma}) b_{RL}(\gamma , \bar{\gamma}) b_{RL}(\beta , \bar{\beta})}{a_{RL} (\gamma , \beta) a_{RL} (\bar{\gamma} , \bar{\beta})} \frac{\delta (\bar{\beta})}{\Delta (\gamma)} \ .
\end{displaymath}
The sum over partitions does not seem to have the correct form to apply (\ref{corollary}), as one set of $a_{RL}$ factors is not ordered in the right way. Furthermore, (\ref{corollary}) assumes that we are dealing with anti-symmetric functions, which is not true here due to the presence of $a_{RL}$. We can circumvent this assumption by applying a more general version of that formula, stated in equation (\ref{corollary2}), together with
\begin{equation}
	\frac{a_{RL} (u_1 -v_j) a_{RL} (u_i - v_1)}{b_{RL} (u_1 -v_j) b_{RL} (u_i - v_1)} - \frac{a_{RL} (u_1 -v_1) a_{RL} (u_i - v_j)}{b_{RL} (u_1 -v_1) b_{RL} (u_i - v_j)}=\frac{-1}{b_{RL} (u_1 - u_i ) b_{RL} (v_1 - v_j)} \ ,
\end{equation}
to get
\begin{equation}
	S( \u | \v)= \frac{ Y^< (\u - \u ) Y^> (\v - \v )}{d^< (\u - \u ) d^> (\v - \v )} \det \left[ \frac{b_{RL} (u-v)}{a_{RL} (u-v)} \left[ 1- \delta (v) \Delta (u) \right] \right] \ . \label{offshellscalarproductmixedfluxRL}
\end{equation}
As in the previous section, we can take out the $d$ functions and write
\begin{equation}
	\langle 0 | \tilde{B}(u_1|\vec{\theta}\, ) \dots \tilde{B}(u_M|\vec{\theta}\, ) B(v_M|\vec{\theta}\, ) \dots B(v_1 |\vec{\theta}\, )|0\rangle= Y^< (\u - \u ) Y^> (\v - \v ) \det \left[ \frac{b_{RL} (u-v)}{a_{RL} (u-v)} \left[ 1-\delta (v) \Delta (u) \right] \right] \ .
\end{equation}

\section{Conclusions}

In this paper, we have taken the first step towards the computation of correlation functions in the massless sector of $AdS_3$ integrable superstring theories, by computing the scalar products and the norms of the Bethe vectors. We have derived determinant-like formulas for the off-shell scalar-products, and we have tested our formulas against explicit calculations involving off-shell Bethe vectors obtained from monodromy matrices of small length, namely one and two frame particles, observing matching. We have computed some simple norms for both off-shell and on-shell vectors in the R-R case, and verified the norms are real and positive. 

In doing this, we have extended the applicability of the determinant-like reformulation to more general functional forms and $R$-matrix settings as it had been done before. We remark that our $S$-matrices are not conventional and purely non-perturbative in nature.

We have pushed the numerical tests for three and four excitations and a general number of frame particles to the limit of our computer capabilities, and we remark that the numerical plot display some difference in some region of the parameters w.r.t. the exact formulas, whereas other regions show a remarkable matching. We do believe that the numerics might be reaching a stage of non-reliability where the curves start differing, however a more thorough numerical check is paramount and we plan to address this issue in the future. 

Thanks to the new difference-form change of variables worked out in \cite{AndreaAle,Fontanella:2019ury} for pure R-R, the pure R-R analysis of scalar products in this paper applies directly to the complete massless sector away from the BMN limit, simply adapting the rapidities we use here to the $\gamma$ variables introduced in \cite{AndreaAle}.

The next stage towards the computation of correlations functions goes via the derivation of form factors for the CFTs described by these $S$-matrices, using the formulas we have obtained here and the quantum inverse scattering problem. Regarding this second point, the construction of local operators (in this case, matrix elements $E_{ij}$ acting on a particular site of the spin chain) from the monodromy matrix would be just a supersymmetric version of the arguments from \cite{Maillet:1999re}, as all the $R$-matrices we have studied here become graded permutation when the argument vanishes. \footnote{Exception to this is (\ref{RmatrixLL}) at $k=1$ (and of course the $R$-matrix scattering different representations (\ref{RmatrixRL}), which is however not expected to). These na\"ive $k=1$ $R$-matrices are actually not the correct ones due to some subtleties involved in the limit, the correct ones being equivalent to their respective $k=2$ cases. We refer to \cite{Fontanella:2019ury} for a complete explanation of the topic.} Nevertheless, this is clearly a challenging task for the very nature of massless integrable theories, and we believe that this will be an ideal setting to address this difficult problem, given the relative simplicity of the $S$-matrix and the exact expressions we have obtained. We leave this task for future work. 

\section{Acknowledgments}
We very much thank Bogdan Stefa\'nski for discussions and for a careful reading of the manuscript. We thank  Marius de Leeuw and Tristan McLoughlin for discussions.
This work is supported by the EPSRC-SFI grant EP/S020888/1 {\it Solving Spins and Strings}. 

No data beyond those presented and cited in this work are needed to validate this study.

\appendix

\section{Generalised versions of Cauchy determinant formula}

\label{trigCauchy}

In the course of this article we have used generalisations of the usual Cauchy determinant formula
\begin{equation}
	\det \left[ \left( \frac{1}{x_i - y_j} \right)_{1\leq i,j \leq M} \right] =\frac{\prod_{1\leq i<j \leq M} (x_i - x_j) (y_j - y_i)}{\prod_{1\leq i,j \leq M} (x_i - y_j)}
\end{equation}
to hyperbolic functions. The proof of all these identities is very similar, so we give here a general proof valid for all our cases of interest.

Given a general function $m(x,y)$ and a matrix of the form ${\cal M}_{i,j}=m(u_i,v_j)$, we can divide each row by the element from the first column and subtract the first row from all other rows. Thus, elements that are neither in the first column nor the first row are of the form
\begin{displaymath}
	\frac{m(u_i,v_j)}{m(u_i,v_1)}-\frac{m(u_1,v_j)}{m(u_1,v_1)}=\left( \frac{1}{m(u_1,v_j) m(u_i,v_1)}-\frac{1}{m(u_i,v_j) m(u_1,v_1)} \right) m(u_1,v_j) m(u_i,v_j) \ .
\end{displaymath}
In all our cases the following identity holds\footnote{This identity not only holds for rational or hyperbolic functions but also for elliptic functions, as Fay's trisecant identity can be written in this form.}
\begin{equation}
\label{CauchyProperty}
	\left( \frac{1}{m(u_1,v_j) m(u_i,v_1)}-\frac{1}{m(u_i,v_j) m(u_1,v_1)} \right)= \epsilon \, g (u_1 , u_i) g(v_1 , v_j) \ ,
\end{equation}
where $\epsilon$ is either $\pm 1$. As both functions $g$ depend either on the row and on the column but not on both, we can take them out of the determinant. We can do similarly with the factor $m(u_1,v_j)$. Thus, we get the determinant of the same matrix but with the first column replaced by ${\cal M}_{1,1}=1$ and ${\cal M}_{i,1}=0$, i.e.
\begin{multline}
	\det_{M\times M} \left[ \left( {\cal M}_{i,j} \right)_{1\leq i,j \leq M} \right] =\\
	= \epsilon^{M-1} m(u_1 , v_1) \left[\prod_{i\neq 1} g (u_1 , u_i) g(v_1 , v_i) m(u_i,v_1) m(u_1,v_i) \right] \det_{(M-1) \times (M-1)} \left[ \left( {\cal M}_{i,j} \right)_{2\leq i,j \leq M} \right] \ .
\end{multline}
This give us a recurrence relation that can be used to perform a proof by induction. Furthermore, (\ref{CauchyProperty}) is proportional to the determinant involving two excitations, providing an initial condition for the proof by induction. The final expression is then
\begin{equation} \label{generalizedcauchy}
	\det_{M\times M} \left[ \left( {\cal M}_{i,j} \right)_{1\leq i,j \leq M} \right] =\epsilon^{\frac{M (M-1)}{2}} \left[ \prod_{i < j} g (u_i , u_j) g(v_i , v_j) \right]  \left[ \prod_{i,j=1}^M m(u_i,v_j) \right]\ .
\end{equation}
Notice that the set of $u$'s and $v$'s are required to have the same cardinality for $\cal M$ to be a square matrix, allowing us to write the factor involving the $g$ functions as either $\prod_{i < j} g (u_i , u_j) g(v_i , v_j)$ or $\prod_{i < j} g (u_j , u_i) g(v_j , v_i)$. Furthermore, if $\epsilon=-1$ we can incorporate this factor into the product and write it as $\prod_{i < j} g (u_i , u_j) g(v_j , v_i)$ or $\prod_{i < j} g (u_j , u_i) g(v_i , v_j)$.

A useful corollary to (\ref{generalizedcauchy}), assuming that $g$ is antisymmetric, is the identity
\begin{equation}
	\sum_{\substack{\{ u_I \} \cup \{ u_{II} \}=\{ \u \} \\ \{ v_I \} \cup \{ v_{II} \}=\{ \v \}}} (-\epsilon)^{|u_I| |u_{II}|} \frac{m(u_I , v_I) m(u_{II} , v_{II}) p(u_I) q(v_{II})}{g(u_{II} , u_I) g(v_I , v_{II})}=(-\epsilon)^{\frac{M(M-1)}{2}} \frac{\det \left[ \vphantom{\frac{a}{b}} m(u_i,v_j) \left[ q(v_i) + p(u_j) \right] \right]}{\left[ \prod_{i < j} g (u_i , u_j) g(v_j , v_i) \right]} \ . \label{corollary2}
\end{equation}
The proof goes as follows: we use the generalised Cauchy determinant formula to rewrite the numerator as the product of the determinant of $m(u_i,v_j) p(u_i)$ over the first set of variables and  the determinant of $m(u_i,v_j) q(v_j)$ over the second set of variables, which can be combined into one determinant of a block-diagonal matrix over all variables. The rows and columns of this matrix are not correlated with the indices of the $u$'s and $v$'s, and reordering them inside the determinant will give us extra signs. However, these signs (together with the $-\epsilon$ factors) can be cancelled with the signs needed to reorder the functions in the denominator in such a way that we have a factor $g^< (\u,\u) g^> (\v,\v)$ in each term of the sum over partitions. Now the sum over partitions becomes the expansion of $m(u_i,v_j) \left[ p(u_i) +q(v_j) \right]$ into a sum of matrices\footnote{Here we have used that the determinant is a multilinear function. This means that if we decompose one the columns of a matrix as two vector columns $\vec{v}+\vec{w}$, then the determinant of this matrix is the sum of the determinants of the matrices obtained from it by replacing such column by $\vec{v}$ and then by $\vec{w}$. A similar property holds also for the rows.} having just either the factor $q(v_i)$ or $p(u_j)$, completing the proof.

Usually it will also be the case that the function $m$ is antisymmetric and $p=q$, allowing us to write the following weaker version of the corollary
\begin{equation}
	(-1)^M \sum_{\text{part.}} (-\epsilon)^{|u_I| |u_{II}|} \frac{m(u_I , v_I) m(v_{II} , u_{II}) h(u_I) h(v_{II})}{g(u_{II} , u_I) g(v_I , v_{II})}=(-\epsilon)^{\frac{M(M-1)}{2}} \frac{\det \left[ \vphantom{\frac{a}{b}} m(u_i,v_j) \left[ h(v_i) - h(u_j) \right] \right]}{\left[ \prod_{i < j} g (u_i , u_j) g(v_j , v_i) \right]} \ . \label{corollary}
\end{equation}

We would like to finish this section writing the explicit form of the different functions $m$ and $g$ we used in this article\footnote{In some cases extra numerical factors or a relabelling of variables might be necessary.}
\begin{enumerate}
	\item The usual Cauchy determinant formula is obtained from this proof by setting $m(u,v)=g(u,v)=\frac{1}{u-v}$ and $\epsilon=-1$.

	\item Equation (\ref{CauchySech}) is obtained from this proof by setting $m(u,v)=g(u,v)=\sech \left( \frac{u-v}{2} \right)$, $\epsilon=1$ and $g(u,v)=\sinh \left( \frac{u-v}{2} \right)$.
	
	\item Equation (\ref{offshellscalarproductnoflux}) is obtained from the corollary by setting $m(u,v)=\frac{1}{g(u,v)}=\csch \left( \frac{u-v}{2} \right)$ and $\epsilon=-1$.
	
	\item Equations (\ref{highestweightmixed}) and (\ref{offshellscalarproductmixedfluxRL}) are obtained form the proof and the stronger corollary respectively by setting $m(u,v)= \frac{X(u-v)}{Y(u-v)}$, $\epsilon=-1$ and $g(u,v)=\frac{1}{X(u-v)}$.
	
	\item Equations (\ref{highestweightmixedRL}) and (\ref{offshellscalarproductmixedflux}) are obtained form the proof and the corollary respectively by setting $m(u,v)=\frac{1}{g(u,v)}= X(u-v)$ and $\epsilon=-1$.
\end{enumerate}

\section{Computation of highest weights}

In this appendix we will explain a method to compute the highest (and lowest) weight for a given number of excitations. The basic premise is to relate the highest weight for the case of $M$ excitations with the one for the case of $M-1$ excitations and solve the resulting recurrence relation, which can be recast as the Laplace expansion of a generalised Cauchy determinant in all the cases of relevance for us.

\subsection{Pure R-R}
\label{highestweights1}

To begin with, we need to compute the action of a single $C$ operator on a stack of $B$ operators. We will not need to know all the terms involved in this expression in order to compute the recurrence relation. In particular, we will only need to know the coefficient appearing in front of the $\prod_{i=1}^{M-1} B(v_i) D(v_M) |0\rangle$ term (assuming that $B(v_M)$ is the first operator from the left), as any other term either do not contribute to the highest weight or can be obtained from the symmetry of $B$ operators. Applying equation (\ref{comCB}) just once we get
\begin{displaymath}
	C(u) \prod_{i=1}^M B(v_i) |0\rangle=\left\{ B(v_M) C(u) + \csch \left( \frac{u-v_M}{2} \right) \left[ D (v_M) A(u) - D (u) A(v_M)\right] \right\} \prod_{i=1}^{M-1} B(v_i) |0\rangle \ .
\end{displaymath}
One can see that the first and third factor are irrelevant for our computation. The first one cannot give a factor of $\prod_{i=1}^{M-1} B(v_i)$ when we try to move the $C$ operator further to the right, while the third one will have the $u$ rapidity either in a $B$ or the $D$ operator when we try to move the $D$ operator to the right. The only relevant contribution will come from the $D (v_M) A(u)$ factor, which has to be moved to the right most part using only the ``wanted terms'' in the commutation relation if we want to get a factor $\prod_{i=1}^{M-1} B(v_i) D(v_M)$. We can see that this is the only possible procedure that gives such factor. Supplementing this result with the symmetry of the $B$ operators, we can easily recover the coefficient accompanying any $\prod_{i\neq j} B(v_i) D(v_j)$
\begin{displaymath}
	C(u) \prod_{i=1}^M B(v_i) |0\rangle=\sum_j \csch\left( \frac{u-v_j}{2} \right)  \prod_{i\neq j} \coth \left( \frac{u-v_i}{2} \right) \coth \left( \frac{v_j-v_i}{2} \right) B(v_i) D(v_j) |0\rangle +\dots \ ,
\end{displaymath}
where the dots represent other terms that do not contribute to the highest weight. A similar reasoning will be applied for the two cases of mixed flux up to the last step, where the non-commutativity of the $B$ operators has to be taken into account.

If we apply this relation to a general scalar product and extract the highest weight from both sides, we get
\begin{equation}
	Z_M (\u | \v)= \sum_j \csch\left( \frac{u_M-v_j}{2} \right)  \prod_{i\neq j} \coth \left( \frac{u_M-v_i}{2} \right) \coth \left( \frac{v_i-v_i}{2} \right) Z_{M-1} ( \bar{u}_M | \bar{v}_j ) \ ,
\end{equation} 
where $\bar{u}_M$ means the set of $u$'s without the rapidity $u_M$, and similarly for $\bar{v}_j$. If we define now an ``undressed highest weight''
\begin{displaymath}
	Z (\u | \v )= \coth^< \left( \frac{\v - \v}{2} \right)  \coth \left( \frac{\u - \v}{2} \right) \tilde{Z} (\u | \v ) \ ,
\end{displaymath}
the recurrence relation simplifies to
\begin{equation}
	\tilde{Z}_M (\u | \v)= \sum_j (-1)^{j-1}\sech \left( \frac{u_M-v_j}{2} \right) \prod_{k\neq M} \tanh \left( \frac{u_k - v_j}{2} \right) \tilde{Z}_{M-1} ( \bar{u}_M | \bar{v}_j ) \ ,
\end{equation}
which is just the Laplace expansion for a determinant, as $(-1)^{j-1}=(-1)^{j+1}$. Combining this formula with the explicit $M=1$ case, given by $\tilde{Z}_1 (u | v)=\sech \left( \frac{u-v}{2} \right)$, we get\footnote{The non-conventional ordering in the $i$ index comes from the Laplace expansions, which tell us to put the rapidity $u_M$ the furthest to the left.}
\begin{equation}
	\tilde{Z}_M (\u | \v)= \det \left[ \sech \left( \frac{u_{M+1-i}-v_{j}}{2} \right) \prod_{k=1}^{M-i} \tanh \left( \frac{u_k - v_j}{2} \right) \right] \ . \label{intermediateHighestWeight}
\end{equation}
We can get rid of the hyperbolic tangent factors by using the fact that the determinant remains unaltered when we add to a column a linear combination of the other columns. If we add to a column the weighted sum of all the columns to their right (with the weight given by the hyperbolic cosecant of the difference of the arguments appearing in their hyperbolic secant, i.e. $\frac{u_k-u_i}{2}=\frac{u_k-v_j}{2}-\frac{u_i-v_j}{2}$) and use the identity
\begin{equation}
	\sech \left( \frac{u_i-v_j}{2} \right)  \tanh \left( \frac{u_k -v_j}{2} \right) +  \csch \left( \frac{u_k-u_i}{2} \right) \sech \left( \frac{u_k-v_j}{2} \right)= \sech \left( \frac{u_i-v_j}{2} \right)  \coth \left( \frac{u_k -u_i}{2} \right) \ ,
\end{equation}
we exchange the hyperbolic tangent factors by hyperbolic cotangent factors that are independent of the column, and thus can be taken out of the determinant. Thus, the final determinant expression for the ``undressed highest weight'' is given by
\begin{equation}
	\tilde{Z}_M (\u | \v)= \coth^< \left( \frac{\u - \u}{2} \right) \det \left[ \sech \left( \frac{u_{M+1-i}-v_{j}}{2} \right) \right] \ .
\end{equation}

Furthermore, the matrix we got looks like a hyperbolic version of a Cauchy matrix, and actually a similar formula holds for it, as explained in the previous appendix. In particular, rearranging the columns to get the same expression as equation (\ref{generalizedcauchy}) at the cost of a factor of $(-1)^{\frac{|M|(|M|-1)}{2}}$, we get
\begin{equation}
	\det \left[ \sech \left( \frac{u_{M+1-i}-v_{j}}{2} \right) \right]= (-1)^{\frac{M(M-1)}{2}} \sinh^< \left( \frac{\u-\u}{2} \right)  \sinh^< \left( \frac{\v - \v}{2} \right)  \sech \left( \frac{\u - \v}{2} \right) \ , \label{CauchySech}
\end{equation}
thus
\begin{equation}
	Z_{|u|} (\u | \v)= (-1)^{\frac{|u|(|u|-1)}{2}} \cosh^< \left( \frac{\u-\u}{2} \right)  \cosh^< \left( \frac{\v - \v}{2} \right)  \csch \left( \frac{\u - \v}{2} \right) \ .
\end{equation}

The same steps can be done for the lowest weight $\bar{Z}$, giving us the same formula but with $u$'s and $v$'s interchanged
\begin{align}
	\bar{Z}_{|u|} (\u | \v) &= (-1)^{\frac{|u|(|u|-1)}{2}} \cosh^< \left( \frac{\u-\u}{2} \right)  \cosh^< \left( \frac{\v - \v}{2} \right)  \csch \left( \frac{\v - \u}{2} \right) \ , \\
	&=(-1)^{|u|+\frac{|u|(|u|-1)}{2}} \cosh^< \left( \frac{\u-\u}{2} \right)  \cosh^< \left( \frac{\v - \v}{2} \right)  \csch \left( \frac{\u - \v}{2} \right) \ ,
\end{align}
where we have used that $|u|=|v|$.

\subsection{Mixed flux, LL-LL case}
\label{highestweights2}

The computation of the highest weight for the case of mixed flux is very similar to the case of pure R-R flux with one important caveat: the B operators no longer commute, which forces us to include factors of $[-c_{LL}(v_k - v_j)]$ in the recurrence relation. We will deal with these extra factors by eliminating the $c_{LL}$ factors through the functions $Y$ involved in the commutation relations and the sign through reordering the Laplace expansion. 

Following the same argument as above, the commutation relations of the $C$ operators and the $B$ operators (\ref{CBcommutationLL}) provide us the recurrence relation
\begin{equation}
	Z_M ( \u |\v )=\sum_{j=1}^M X(u_M - v_j) \prod_{k\neq j} Y(u_m - v_k) Y(v_j - v_k) \prod_{l>j} [-c_{LL}(v_k - v_j)] Z_{M-1} (\bar{u}_M | \bar{v}_j) \ ,
\end{equation}
where $Z_M ( \u |\v )$ is defined as the highest weight for $\langle \overrightarrow{\prod_i} C(u_i) \overleftarrow{\prod_i} B(v_i) \rangle$, \emph{without} including the $d$ functions that symmetrise the states. This recursion relation can be related to a Laplace expansion of a determinant, which can be seen very clearly if the rewriting $\tilde{Z}_M ( \u |\v )=\frac{Z_M ( \u |\v )}{Y(\u - \v) Y^> (\v - \v)}$ is performed\footnote{Notice that, in contrast with (\ref{intermediateHighestWeight}), the product goes up to $i-1$ instead of $M-i$. This is not because the structure of the product is different, but due to the ordering of the index $i$ inside the matrix imposed by the extra minus signs.}
\begin{equation}
	\tilde{Z}_M ( \u |\v )=\det \left[ \frac{X (u_i - v_j)}{Y(u_i - v_j)} \prod_{k=1}^{i-1} \frac{1}{Y(u_k - v_j)} \right] \ .
\end{equation}
This expression for the highest weight can be further simplified if we use the following two properties of the $X$ and $Y$ functions
\begin{align}
	\frac{X (u_i - v_j)}{Y(u_i - v_j)} \frac{1}{Y(u_k - v_j)} + X(u_k - u_i) \frac{X(u_k - v_j)}{Y(u_k - v_j)} &= Y(u_i - u_k) \frac{X(u_i - v_j)}{Y(u_i - v_j)} \ , \\
	\frac{Y (u_1 - v_j) Y(u_i - v_1)}{X (u_1 - v_j) X(u_i - v_1)}-\frac{Y (u_1 - v_1) Y(u_i - v_j)}{X (u_1 - v_1) X(u_i - v_j)} &= \frac{i}{X(u_1 - u_i)} \frac{i}{X(v_1 - v_j)} \ .
\end{align}
The first one allows us to replace the product of $\frac{1}{Y}$ functions appearing inside the matrix by a global factor $Y^> (\u - \u )$ while the second one gives us the necessary condition to apply the generalisation of the Cauchy determinant formula (\ref{generalizedcauchy}). Thus, the final form of the highest weight is
\begin{equation}
	Z_{|u|} (\u | \v) = (-1)^{\frac{|u|(|u|-1)}{2}} \frac{Y^> (\u - \u) Y^> (\v - \v )}{X^> (\u - \u) X^> (\v - \v )} X(\u - \v ) \ . \label{highestweightmixed}
\end{equation}

\subsection{Mixed flux, RL-LL case}
\label{highestweights3}

From the commutation relations (\ref{commutationRLwithLL}) we can compute the contribution $\prod_{i\neq M} B(v_i) D(v_M) |0\rangle$ in $\tilde{B} (u_M) B(\v) |0\rangle$ and use it to construct the recurrence relation between highest weights with different number of excitations following the arguments explained above. In particular, we get
\begin{equation}
	\hat{Z}_M (\u | \v) =\sum_{j=1}^M \frac{-b_{RL} (u_M -v_j)}{a_{RL} (u_M -v_j)} \prod_{k\neq j} \frac{Y(v_j -v_k)}{a_{RL} (u_m -v_k)} \prod_{l>j} \left[-c(v_k - v_j) \right] \hat{Z}_{M-1} (\bar{u}_M | \bar{v}_j) \ .
\end{equation}
We can get a Laplace expansion of a determinant if we take out a factor $(-1)^M \frac{a_{RL} (\u - \v)}{Y^> (\v - \v)}$, and use it reconstruct the following intermediate expression
\begin{equation}
	\hat{Z}_M (\u | \v)= (-1)^M \frac{a_{RL} (\u - \v)}{Y^> (\v - \v)} \det \left[ b_{RL} (u_i - v_j) \prod_{l=1}^{i-1} a_{RL} (u_k - v_j) \right] \ .
\end{equation}
This expression can be simplified if we make use of the relations
\begin{align*}
	b_{RL} (u_i - v_j) a_{RL} (u_k - v_j) + i b_{RL} (u_i - u_k) b_{RL} (u_k -v_j) = a_{RL} (u_k - u_i) b_{RL}(u_i - v_j) \ , \\
	\frac{1}{b_{RL} (u_1 , v_j) b_{RL} (u_i , v_1)} - \frac{1}{b_{RL} (u_1 , v_1) b_{RL} (u_i , v_j)} = \frac{-1}{b_{RL} (u_1 - u_i ) b_{RL} (v_1 - v_j)} \ ,
\end{align*}
and take into account that $a_{RL} (x)$ and $Y(x)$ are actually the same function. 
Using the first equation to get rid of the extra $a_{RL}$ factors in the determinant and the second one to apply (\ref{generalizedcauchy}), we can write the highest and lowest weights as
\begin{equation}
	\hat{Z}_{|u|} (\u | \v) = (-1)^{|u|} \hat{\bar{Z}}_{|u|} (\u | \v)= (-1)^{|u|} \frac{ b_{RL} (\u - \v) a_{RL}^< (\u - \u ) a_{RL}^> (\v - \v )}{ a_{RL} (\u - \v) b_{RL}^< (\u - \u ) b_{RL}^> (\v - \v )} \ . \label{highestweightmixedRL}
\end{equation}

\section{\label{new} On-shell norms in the pure R-R case}

In this appendix we are going to use brute force and the method described in \cite{JuanMiguel} to compute  off-shell norms for states with a low number of magnons and a general on-shell norm for any number of magnons, both in the pure R-R case. As described in section \ref{Conj}, we have to distinguish here between symmetric or asymmetric norms depending on the imaginary part of the inhomogeneities $\theta_i$, $i=1,...,L$, corresponding to the rapidity of the so-called {\it frame} particles (see section \ref{ABA}). We will use the subindex $(1)$ for the norms computed assuming real inhomogeneities and the subindex $(2)$ for the ones computed using inhomogeneities with Im$[\theta_j]=-i \pi/2$.

We shall first compute the norm for off-shell Bethe vectors, namely, without imposing the Bethe equations, for some simple cases, and only afterwards we shall impose the Bethe equations quantising the rapidities $u_i$. With that information we will derive the equivalent of the Gaudin determinant for this spin chain. We will also occasionally restrict to the case where all the inhomogeneities are set to be equal.

The computation of the norm for a low number of magnons can be done by use of the commutation relations between the operators $B$ and $C$, which in turn are dictated by the RTT relations described in section \ref{ABA}. We report here the ones which intervene in the calculation:
\begin{eqnarray}
&&C(u_1)B(u_2) =B(u_2)C(u_1) + x(u_1-u_2) D(u_2)A(u_1)-x(u_1-u_2) D(u_1)A(u_2) \ , \\
&&C(u_1)A(u_2) =y(u_1-u_2) A(u_2)C(u_1)-x(u_1-u_2) A(u_1)C(u_2) \ , \\
&&C(u_1)D(u_2) =y(u_1-u_2) D(u_2)C(u_1)-x(u_1-u_2) D(u_1)C(u_2) \ ,
\end{eqnarray} 
where
\begin{equation}
x(u_i-u_j) = \frac{a(u_i - u_j)}{b(u_i - u_j)} = \csch \frac{u_i - u_j}{2}, \qquad y(u_i-u_j) = \frac{1}{b(u_i - u_j)}= \coth\frac{u_i - u_j}{2} \ .
\end{equation} 
By repeated use of these commutation relations, we can bring all the $C$'s near the pseudovacuum $|0\rangle$, which is the state made of all bosons in each quantum space and is annihilated by any $C(u)$, until no $C$'s and $B$'s are left.

\subsection{One excitation} 

A simple computation shows that
\begin{equation}
\label{si}
	\langle 0| C(u) B(v)|0\rangle =x (v-u) \Big[\prod_{i=1}^L b(u - \theta_i) - \prod_{i=1}^L b(v - \theta_i) \Big] \ .
\end{equation}
It is easy to see that the norm of the one magnon state in the asymmetric prescription is given by
\begin{align}
\langle 0| B^\dagger(u) B(u)|0\rangle_{(1)} &= -i \left[ \prod_{i=1}^L \coth \left( \frac{u - \theta_i + i\pi}{2} \right) \right] \csch \left( \frac{-i\pi}{2} \right) \times \notag\\
&\qquad \qquad \qquad \times \left[ \prod_{i=1}^L \tanh \left( \frac{u - \theta_i + i\pi}{2} \right)- \prod_{i=1}^L \tanh \left( \frac{u - \theta_i}{2} \right) \right]  \notag \\
& = 1 - \prod_{i=1}^L b^2(u - \theta_i) = 1 - \prod_{i=1}^L \tanh^2 \frac{u - \theta_i}{2} \ .
\end{align} 
The norm is manifestly real and positive, as all the rapidities have real values. Interestingly enough, the norm is proportional to the auxiliary Bethe equations (\ref{auxi}).

In the symmetric case we have to resolve a $\frac{0}{0}$ ambiguity coming from (\ref{si}), which we do by simply calculating the limit
\begin{equation}
\langle 0| C(u) B(u)|0\rangle \equiv \lim_{v \to u} \langle 0| C(u) B(v)|0\rangle = - 2 \frac{\partial}{\partial u} \prod_{i=1}^L b(u - \theta_i) \ . 
\end{equation}
This gives 
\begin{align}
	\langle 0| B^\dagger(u) B(u)|0\rangle_{(2)} &= 2 i \left[ \prod_{i=1}^L \coth \left( \frac{u - {\rm Re}[\theta_i] + i\pi/2}{2} \right) \right] \frac{\partial \prod_{i=1}^L b(u - {\rm Re}[\theta_i] + i \pi /2)}{\partial u} \notag \\
	&=2 \sum_{i=1}^L \sech (u - {\rm Re}[\theta_i]) \ ,
\end{align} 
which is again manifestly real and positive. 
 
\subsection{Two excitations} 

The $M=2$ case is slightly more involved, and requires the help of computer algebra to keep track of all the terms arising from the commutators. The calculation consists of evaluating
\begin{equation}
\label{return}
\langle 0| B^\dagger(u_1) B^\dagger(u_2) B(u_2) B(u_1) |0\rangle \ .
\end{equation} 
For two sites ($L=2$), for instance, this produces, in the asymmetric case, 
\begin{eqnarray}
&&\langle 0| B^\dagger(u_1) B^\dagger(u_2) B(u_2) B(u_1) |0\rangle_{(1)}=\frac{16 e^{u_1 + u_2 + \theta_1 + \theta_2} \, (e^{u_1} + e^{u_2})^2 \, (e^{\theta_1} + e^{\theta_2})^2}{(e^{u_1} + e^{\theta_1})^2 \,(e^{u_1} + e^{\theta_2})^2 \,(e^{u_2} + e^{\theta_1})^2 \,(e^{u_2} + e^{\theta_2})^2}=\nonumber\\
&& \qquad \qquad = \Big[1 - \prod_{i=1}^2 b^2(u_1 - \theta_i)\Big] \Big[1 - \prod_{i=1}^2 b^2(u_2 - \theta_i)\Big]\frac{a(2 u_1 - \theta_1 - \theta_2) a(2 u_2 - \theta_1 - \theta_2)}{a^2(u_1 - u_2)} \ ,
\end{eqnarray}
where we remind that $a(x) = {\rm sech} \frac{x}{2}$ and $b(x) = \tanh \frac{x}{2}$.
The norm is manifestly real and positive, since all the rapidities are real, and it is again proportional to the auxiliary Bethe equations.

For the symmetric case we have
\begin{align}
\label{C}
	&\langle 0| C(u_1) C(u_2) B(u_2) B(u_1) )|0\rangle_{(2)} =  \notag\\
	&  \qquad \qquad \qquad \, \, \, 4\coth^2 \left( \frac{u_1 - u_2}{2} \right) D_1 D_2 \times \prod_{j=1}^L b(u_1 - {\rm Re}[\theta_j]+i\pi/2) b(u_2 - {\rm Re}[\theta_j]+i\pi/2) \notag \\
	&-\frac{1}{4} \csch^6 \left( \frac{u_1 -u_2}{2} \right) \sinh^2 (u_1 -u_2) \left[ \prod_{i=1}^L b(u_1 - {\rm Re}[\theta_i]+i\pi/2) - \prod_{i=1}^L b(u_2 - {\rm Re}[\theta_i]+i\pi/2) \right]^2 \ ,
\end{align} 
where $D_j$ stands for the sum of the logarithmic derivatives of the function $b$ w.r.t. the auxiliary variable over all the inhomogeneities, {\it i.e.},
\begin{equation}
	D_j = \sum_{k=1}^L \frac{1}{b (u_j -{\rm Re}[\theta_k]+i\pi/2)} \frac{\partial b (u_j -{\rm Re}[\theta_k]+i\pi/2)}{\partial u_j} \ = - i \sum_{k=1}^L {\rm sech}(u_j -{\rm Re}[\theta_k]).
\end{equation}
Notice that the term involving no derivatives is proportional to the two-magnons Bethe equations, thus it will vanish when they are imposed. This means that, upon restricting to on-shell Bethe vectors, we can write the expression for the norm in the symmetric case as
\begin{equation}
\langle 0| B^\dagger(u_1) B^\dagger(u_2) B(u_2) B(u_1) )|0\rangle_{(2)} = 4 \coth^2 \left( \frac{u_1 - u_2}{2} \right) \sum_{i=1}^L \sech (u_1 - {\rm Re}[\theta_i]) \sum_{i=1}^L \sech (u_2 - {\rm Re}[\theta_i]) \ ,\nonumber
\end{equation} 
where we have used the auxiliary Bethe equations to set to $1$ the term $\prod_{j=1}^L b(u_1 - {\rm Re}[\theta_j]+i\pi/2) b(u_2 - {\rm Re}[\theta_j]+i\pi/2)$ in (\ref{C}), and to simplify the factor appearing in (\ref{multi}) when transitioning from (\ref{C}) to the actual norm.

We have to remark that, in the case of all inhomogeneities being equal, the norm only has a $L^2$ contribution but no term linear in $L$. This can be understood from the structure of the usual Gaudin determinant, where the Gaudin matrix is proportional to the derivative of the $i$-th Bethe equation with respect to the $j$-th rapidity. In this case the $i$-th Bethe equation only depends on the $i$-th rapidity as there is no $S$-matrix term involved, which implies that the Gaudin determinant is diagonal.

\subsection{General number of excitations} 

Let us construct now the analogue of the Gaudin norm, i.e. the on-shell norm in the asymmetric case, for a state with a general number of magnons using the composite model trick explained in section \ref{Recap} and expanding in powers of the length of the chain. From the cases of one and two magnons we find that
\begin{equation}
	\sigma^{(1)}_1 (u) = 2 \sum_{i=1}^L \sech (u - {\rm Re}[\theta_i]) \ ,  \qquad \sigma^{(2)}_1=0 \ .
\end{equation}
The separability hypothesis holds in this case, namely $\sigma^{(l)}_1\propto \prod \sigma^{(2)}_1$, then the only contribution to the norm is the term
\begin{multline}
	\langle 0 |\prod_{j=1}^M C(u_j) \prod_{j=1}^M B(v_j) |0\rangle = f^< (\u , \u ) f^> (\u , \u ) \sigma_1^{(1)} (\{ \beta\}) \\
	=  2^M \prod_{i< j} \coth^2 \left( \frac{u_i - u_j}{2} \right) \prod_{j=1}^M \sum_{k=1}^L \sech (u_j - {\rm Re}[\theta_k]) \ .
\end{multline}
Notice that the norm is once again manifestly real and positive. We can see that this result matches the norm limit of equation (\ref{offshellscalarproductnoflux}) when we impose the Bethe ansatz equations.

\end{document}